\documentclass[journal,onecolumn,11pt]{IEEEtran}
\usepackage[T1]{fontenc}
\usepackage[utf8]{inputenc}
\usepackage{cite}
\usepackage{amsmath,amssymb,amsthm,mathtools,bm}
\usepackage{booktabs,array,tabularx,microtype}
\usepackage{graphicx,aliascnt,enumitem}
\usepackage[hidelinks,pdfencoding=auto,psdextra]{hyperref}
\usepackage[capitalise,nameinlink,noabbrev]{cleveref}
\input{glyphtounicode}
\newcolumntype{Y}{>{\raggedright\arraybackslash}X}
\newtheorem{theorem}{Theorem}
\newaliascnt{lemma}{theorem}
\newtheorem{lemma}[lemma]{Lemma}
\aliascntresetthe{lemma}
\newaliascnt{proposition}{theorem}
\newtheorem{proposition}[proposition]{Proposition}
\aliascntresetthe{proposition}
\newaliascnt{corollary}{theorem}
\newtheorem{corollary}[corollary]{Corollary}
\aliascntresetthe{corollary}
\theoremstyle{remark}
\newaliascnt{remark}{theorem}
\newtheorem{remark}[remark]{Remark}
\aliascntresetthe{remark}
\theoremstyle{definition}
\newaliascnt{definition}{theorem}
\newtheorem{definition}[definition]{Definition}
\aliascntresetthe{definition}
\crefname{theorem}{Theorem}{Theorems}
\Crefname{theorem}{Theorem}{Theorems}
\crefname{lemma}{Lemma}{Lemmas}
\Crefname{lemma}{Lemma}{Lemmas}
\crefname{proposition}{Proposition}{Propositions}
\Crefname{proposition}{Proposition}{Propositions}
\crefname{corollary}{Corollary}{Corollaries}
\Crefname{corollary}{Corollary}{Corollaries}
\crefname{remark}{Remark}{Remarks}
\Crefname{remark}{Remark}{Remarks}
\crefname{definition}{Definition}{Definitions}
\Crefname{definition}{Definition}{Definitions}
\newcommand{\bZ}{\bm Z}

\newcommand{\ind}{\perp\!\!\!\perp}
\newcommand{\TC}{\operatorname{TC}}
\newcommand{\PaperAuthorLine}{Han Fang, Nan Liu, and Wei Kang}
\newcommand{\PaperAuthorFootnotes}{%
\thanks{Han Fang is with the School of Cyber Science and Engineering,
Southeast University, Nanjing, China (e-mail: h\_fang@seu.edu.cn).}%
\thanks{Nan Liu is with the National Mobile Communications Research Laboratory
and the School of Cyber Science and Engineering, Southeast University,
Nanjing, China (e-mail: nanliu@seu.edu.cn).}%
\thanks{Wei Kang is with the School of Information Science and Engineering
and the School of Cyber Science and Engineering, Southeast University,
Nanjing, China (e-mail: wkang@seu.edu.cn).}}
\newcommand{\PaperPDFAuthors}{Han Fang, Nan Liu, Wei Kang}

\title{The Exact Memory--Rate Tradeoff for Three-User Secure Coded Caching}
\author{\PaperAuthorLine\PaperAuthorFootnotes}
\date{}
\hypersetup{pdftitle={The Exact Memory-Rate Tradeoff for Three-User Secure Coded Caching},pdfauthor={\PaperPDFAuthors}}
\begin{document}
\maketitle
\begin{abstract}
We determine the exact zero-error memory--rate tradeoff for secure
coded caching with three users and any number $N\ge2$ of independent
files. Each user recovers its requested file, while its cache and the
broadcast jointly reveal no information about the complete tuple of
unrequested files. For $N\ge4$, two converse inequalities establish the
optimality of the known achievable points and the memory-sharing
segments between them. For $N=3$, two additional bounds identify a new
corner at $(M,R)=(3/2,2)$, where memory and rate are measured in file
units. An explicit linear code over $\mathbb F_3$ attains this corner
and reduces the rate by one quarter of a file relative to memory sharing
between the known points. The converse links the information held by
cache pairs and triples to delivery under repeated and distinct requests.
Its equality conditions guide the construction: users cancel the keys
to recover their requested files, while uniform masking protects the
complete tuple of unrequested files.
\end{abstract}


\section{Introduction}\label{sec:introduction}

Secure coded caching asks how much communication is needed when
each user must recover one file and learn nothing about the others.
A server stores a library of independent files and fills user caches
before the requests are known. Once the requests arrive, the server
sends a common broadcast. Coordinating cache placement and delivery
allows one transmission to serve several users
\cite{maddahali2014fundamental}. File security requires this coordination
to protect the complete collection of files that each user did not
request \cite{ravindrakumar2017private}.
Known schemes attain the minimum-memory and minimum-rate endpoints,
as well as an intermediate coded-placement point
\cite{fang2026secure}. The exact tradeoff between these points has
remained undetermined for three users. We resolve this problem for
every library size $N\ge2$. The known points are optimal for $N\ge4$.
For $N=3$, we find and attain an additional corner below their
memory-sharing curve.

\subsection{Related Work}

Coded caching uses placement to create multicast opportunities before
the requests are known. Maddah-Ali and Niesen
\cite{maddahali2014fundamental} established this shared-link model and
showed that a coded transmission can serve several users with different
requests. Yu et al.\ \cite{yu2018uncoded} determined
the exact average and worst-case rates under uncoded placement. Their
delivery rule removes redundancy caused by repeated requests, and their
converse separates demands according to the multiplicities of the
requested files. For general coded placement, Tian
\cite{tian2018symmetry} formalized file and user symmetries and showed
how the joint entropy structure of tight converses can guide code
design. These ideas provide a basis for studying how one placement
serves different demand patterns.

File security adds a constraint on what a user can learn from its cache
and the broadcast together. Ravindrakumar et al.\ \cite{ravindrakumar2017private} introduced this requirement for the
complete tuple of unrequested files. Their construction encodes each
file into secret shares, places a subset that reveals no file information
in any one cache, and protects multicast messages with keys shared by their
intended recipients. They also derived security-aware cut-set lower
bounds. When the library contains at least as many files as there are
users, their achievable rate is within a factor of 16 of the optimum
throughout the feasible memory range. The construction separates two roles for randomness:
masking the file information stored in caches and protecting the
transmissions used for recovery.
Subsequent work improves the use of demands and the cost of implementing
this secret-sharing approach. Suthan et al.\ \cite{suthan2017common} analyzed when transmission keys can be removed
without exposing unrequested files. Exploiting repeated requests then
reduces the average delivery rate, while the all-distinct demand load
agrees with the original scheme. Meel and Rajan
\cite{meel2021pda} constructed secretive schemes from placement delivery
arrays (PDAs), which specify both the cache pattern and the multicast
groups. The Maddah-Ali--Niesen PDA recovers the original achievable
points, and suitable PDAs reduce the number of packets into which files
must be split. Repeated demands and subpacketization thus provide
distinct ways to improve a given construction.

The closest memory--rate results are due to Fang et al.\ \cite{fang2026secure}. They reused randomness already present in secret
sharing to reduce the additional keys needed for delivery. Their schemes
and converses determine the minimum-memory and unit-rate endpoints for
arbitrary numbers of files and users, as well as a low-memory segment
for two files. For three users and at least three files, their results
give the achievable points $(1,3)$, $((N+1)/2,3/2)$, and $(2N-2,1)$,
with memory and rate measured in file units. The present work proves
that the two intervening memory-sharing segments are optimal for
$N\ge4$ and identifies the additional three-file corner $(3/2,2)$.

Secure coded caching has also been studied when the cache access pattern or
the observing users change. In shared-cache systems, several users
access the same helper cache, while each user has a separate cache for
delivery keys. Meel and Rajan \cite{meel2021shared} developed secretive
schemes for this topology, and Peter et al.\ \cite{peter2022sharedpda} used PDAs to obtain schemes with lower
subpacketization. Here each user has one dedicated cache, whose full
storage contributes to the memory budget. This topology permits the
converse to compare the information revealed by pairs and triples of
user caches.

The observer's access also determines the security constraint. Secure
delivery protects the files from an external observer of the broadcast.
Sengupta et al.\ \cite{sengupta2015securedelivery} combined
cached keys with coded multicast transmissions and established
constant-gap guarantees. Bahrami et al.\ \cite{bahrami2017secureuncoded}
proved the exact secure-delivery tradeoff under uncoded file placement
when the number of files is at least the number of users. In the
secure coded caching model considered here, the observer is a user with access
to its own cache, and the protected information is its unrequested
file tuple. The converse allows arbitrary randomized coded placement.

Demand privacy protects the indices of other users' requested files.
Wan and Caire \cite{wancaire2021privatedemands} developed private-demand
schemes using virtual users and MDS coding. Yan and Tuninetti
\cite{yan2021demandprivacy} introduced privacy keys and converse bounds
under a condition that protects requests against any colluding subset
of users even when the library is known. Gurjarpadhye et al.\ \cite{gurjarpadhye2022demandprivate} developed further constructions
from restricted demand sets and characterized exact regimes,
including the two-user case with at least two files. Gholami et al.\ \cite{gholami2024privatecaches} additionally hid cache metadata and
constructed schemes from private information retrieval protocols.
These models impose privacy on requests or cache descriptions. In
this paper the requests are public, and the secrecy constraint concerns
the file contents.
Joint file and demand privacy protects the unrequested files and the
other users' requests together. Qi and Ravi \cite{qi2022filedemand}
introduced this joint requirement. For distinct requests, multilayer
constructions give additional achievable memory--rate points
\cite{qi2024filedemand,fang2026joint}. For all repeated and
distinct demand vectors, Fang et al.\ \cite{fang2026joint}
gave a general-user scheme and converse with a multiplicative gap of
$9/2$, together with exact endpoints and adjacent optimal regimes.
The present result determines the entire boundary for three users
under file security with public demands. Its new construction attains
the corner selected by the equality conditions of the converse.

\subsection{Results}

Let $M$ be the cache size and $R$ the worst-case delivery rate, both
measured in file units. For $N\ge4$, the exact boundary on
$1\le M\le2N-2$ consists of the two line segments joining
\[
 (1,3),\qquad \left(\frac{N+1}{2},\frac32\right),\qquad (2N-2,1).
\]
We prove the supporting inequalities
\[
 3M+(N-1)R\ge3N,\qquad
 M+(3N-5)R\ge5N-7.
\]
They match the known achievable points and their memory-sharing
segments. All converse bounds apply to arbitrary placement and
delivery functions satisfying the model's recovery and security
conditions. The rate remains equal to one for $M\ge2N-2$.

For $N=3$, we prove
\[
 2M+R\ge5,\qquad M+R\ge\frac72.
\]
The lines intersect at $(3/2,2)$. An explicit code over $\mathbb F_3$
attains this point using two symbols per file, three symbols per cache,
and four broadcast symbols in the worst case. At the same memory,
sharing the known points $(1,3)$ and $(2,3/2)$ gives rate $9/4$.
The new code saves one quarter of a file per worst-case delivery.
The two additional bounds, the general high-memory bound, and the
achievable points determine the entire three-file boundary.
For $N=2$, the boundary is $R=\max\{3-M,1\}$ for $M\ge1$.

\subsection{Proof Ideas}

A user's cache and a broadcast for a nonconstant demand each reveal
no information about the library on their own. Together, they recover
the requested file and thus form two perfect shares of it.
The converse measures how much file information becomes available
when caches are combined. We use a symmetric code, for which file
and user relabelings preserve the relevant entropies.
Write $W_i$ for file $i$, let $F=H(W_i)>0$, and let $Z_k$ be user $k$'s
cache. For $N\ge3$, define
\[
 \begin{aligned}
 p&=I(W_2;Z_1,Z_2\mid W_1),\\
 q&=I(W_3;Z_1,Z_2,Z_3\mid W_1,W_2).
 \end{aligned}
\]
Thus $p$ is the information a cache pair holds about a second file
when one file is given. The quantity $q$ is the corresponding
information about a third file in all three caches.
Let $X_{123}$ denote the broadcast when users 1, 2, and 3 request
files 1, 2, and 3, respectively. Successive decoding gives
\[
 H(X_{123})\ge3F-p-q.
\]
Hence $p+q$ limits the communication saved under distinct requests.
Repeated requests impose a complementary constraint. The observations
of two users recovering the same file overlap in the broadcast.
Adding the third user's observation, which recovers another file,
forces a three-file-unit entropy budget. The broadcast must also
supply the information about that second file missing from the
caches. Placement symmetry gives the same residual uncertainty for
an unrequested third file. These facts yield the three-core
inequality and, together with the cache constraint
$(N-1)p\le H(Z_1)-F$, the low-memory supporting line.

At high memory, we compare collections of broadcasts under which
one user changes its request while the other requests stay fixed.
One cache and these broadcasts recover several files.
Security gives a lower bound on their joint entropy, and
submodularity combines two overlapping recovery collections.
Retaining the conditional entropy of one selected broadcast on both
sides makes it cancel. The remaining broadcast counts give the
coefficient $3N-5$ in the high-memory supporting line.

At the new three-file corner, equality for a finite symmetric code
forces $p=0$ and $q=F$. A pair of caches therefore reveals no second
file when one file is given, while all three caches recover the
third file when the other two are given. We realize this pattern
with a linear placement. For each nonconstant demand, the
key coefficients span a five-dimensional subspace of each user's
seven-coordinate observation. The requested file maps invertibly
onto the two-dimensional quotient by that subspace, and the
unrequested files map to zero. Uniform independent keys make the
observation distribution depend only on the requested file.
The construction uses linear secret sharing and coset coding
\cite{cramer2000,ozarow1984}. A single interpolation rule gives the
file coefficients for all nonconstant demands. For constant demands,
the server broadcasts the requested file.

\Cref{sec:model} states the model and exact boundary.
The local converse bounds lead to the three-file construction in
\Cref{sec:n3ach}. \Cref{sec:low,sec:high} establish the general
supporting inequalities. The appendices give the construction coordinates, further equality
analysis, and a general form of the entropy identity used in the converse.

\section{Model and Exact Boundary}\label{sec:model}

One server holds a library of files and serves three users through a
shared noiseless link. All logarithms are to base two. For a nonnegative
integer $n$, write $[n]=\{1,\ldots,n\}$, with $[0]=\emptyset$.
For an indexed family of random variables $(U_i)$ and an index set
$\mathcal I$, let $U_{\mathcal I}$ denote the tuple with those indices
in increasing order. We also write $U_{a:b}=(U_a,\ldots,U_b)$ when
$a\le b$ and take this tuple to be empty when $a>b$. We use
$H(\emptyset)=0$ throughout.

At a finite blocklength, the library consists of $N\ge2$ mutually
independent files $W_1,\ldots,W_N$, each uniformly distributed on the same
finite alphabet $\mathcal W$. Write $F=H(W_i)=\log_2|\mathcal W|>0$
for their common entropy. The server has private randomness $S$, independent of
the library, so $S\ind W_{[N]}$. For user $k\in[3]$, let $\mathcal Z_k$
be the finite cache alphabet and let $\phi_k$ be its deterministic
placement function. The random cache content is
\[
  Z_k=\phi_k(W_{[N]},S).
\]
Each demand is a fixed, public vector
$\bm d=(d_1,d_2,d_3)\in[N]^3$: user $k$ requests file $W_{d_k}$.
Let $\mathcal X_{\bm d}$ be the finite broadcast alphabet and
$\psi_{\bm d}$ the deterministic delivery function for this demand.
The broadcast for demand $\bm d$ is
\[
  X_{\bm d}=\psi_{\bm d}(W_{[N]},S).
\]
A demand is constant if $d_1=d_2=d_3$ and nonconstant otherwise.
For $a,b,c\in[N]$, we abbreviate $X_{(a,b,c)}$ as $X_{abc}$.
For example, $X_{112}$ corresponds to demand $(1,1,2)$.
When $N\ge3$, $X_{123}$ corresponds to demand $(1,2,3)$.

\begin{figure}[!t]
\centering
\includegraphics[width=0.78\linewidth]{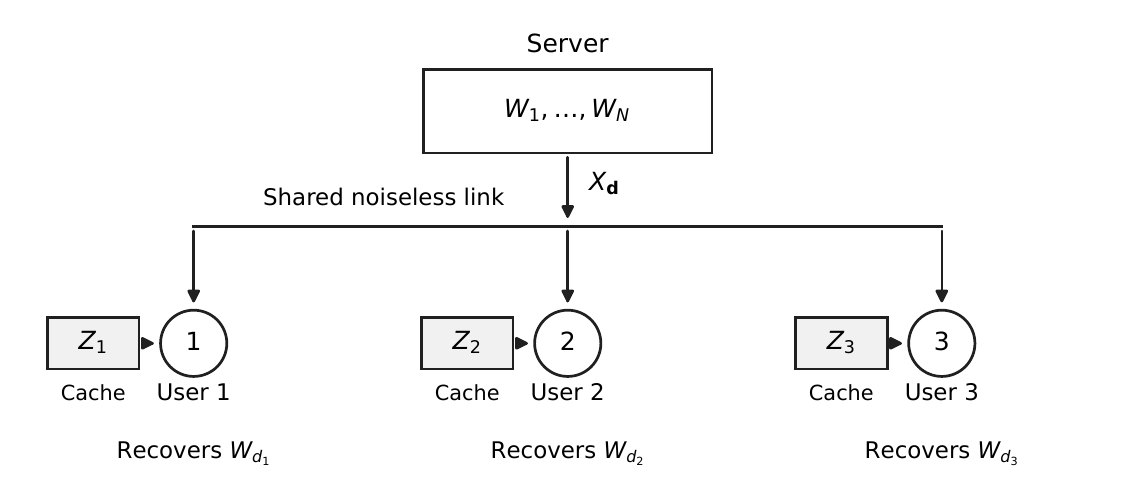}

\caption{Three-user coded caching over the classical shared-link topology
\cite{maddahali2014fundamental,ravindrakumar2017private}. A common broadcast
$X_{\bm d}$ and cache $Z_k$ allow user $k$ to recover $W_{d_k}$.
The file-security requirement is given in \eqref{eq:privacy}.}
\label{fig:system-model}
\end{figure}

For every fixed demand $\bm d$ and every user $k$, the observation
$(Z_k,X_{\bm d})$ must permit zero-error decoding and preserve joint security
of the complete tuple of unrequested files:
\begin{align}
  H(W_{d_k}\mid Z_k,X_{\bm d})&=0,
  \label{eq:decode}\\
  I\bigl(W_{[N]\setminus\{d_k\}};Z_k,X_{\bm d}\bigr)&=0.
  \label{eq:privacy}
\end{align}

The memory budget $M$ and worst-case delivery rate $R$ are normalized by
the file entropy $F$. A memory--rate pair $(M,R)$ is achievable if, for
every $\epsilon>0$, there is a finite-block code satisfying
\eqref{eq:decode}--\eqref{eq:privacy} exactly and
\[
  \frac{\log_2|\mathcal Z_k|}{F}\le M+\epsilon,
  \qquad
  \frac{\log_2|\mathcal X_{\bm d}|}{F}\le R+\epsilon
\]
for every user $k$ and every demand $\bm d$.  Taking the closure permits arbitrary file splitting and memory sharing.  The alphabet constraints imply the entropy bounds
\[
  H(Z_k)\le(M+\epsilon)F,
  \qquad
  H(X_{\bm d})\le(R+\epsilon)F.
\]
We apply the finite-code converse bounds using these entropy upper bounds
and then let $\epsilon\downarrow0$. Let $R^\star_{N,3}(M)$ denote the
infimum achievable worst-case delivery rate at memory budget $M$ for
three users and $N$ files.
All delivery variables $X_{\bm d}$ are functions of the same $(W_{[N]},S)$,
so they are jointly defined across demands.

\subsection{Exact Boundary}

\begin{theorem}[Exact Tradeoff for Three Users]\label{thm:main}
For every $N\ge2$, $R^\star_{N,3}(M)=+\infty$ when $M<1$.  For $M\ge1$ and $N=2$,
\[
  R^\star_{2,3}(M)=\max\{3-M,1\}.
\]
For $M\ge1$ and $N=3$,
\[
  R^\star_{3,3}(M)
  =\max\left\{5-2M,\frac72-M,2-\frac M4,1\right\}.
\]
For every $N\ge4$ and $M\ge1$,
\[
  R^\star_{N,3}(M)
  =\max\left\{
    \frac{3(N-M)}{N-1},
    \frac{5N-7-M}{3N-5},
    1
  \right\}.
\]
For $N\ge4$, the boundary joins $(1,3)$,
$((N+1)/2,3/2)$, and $(2N-2,1)$ and remains at rate one thereafter.
\end{theorem}

\begin{figure}[!t]
\centering
\includegraphics[width=0.98\linewidth]{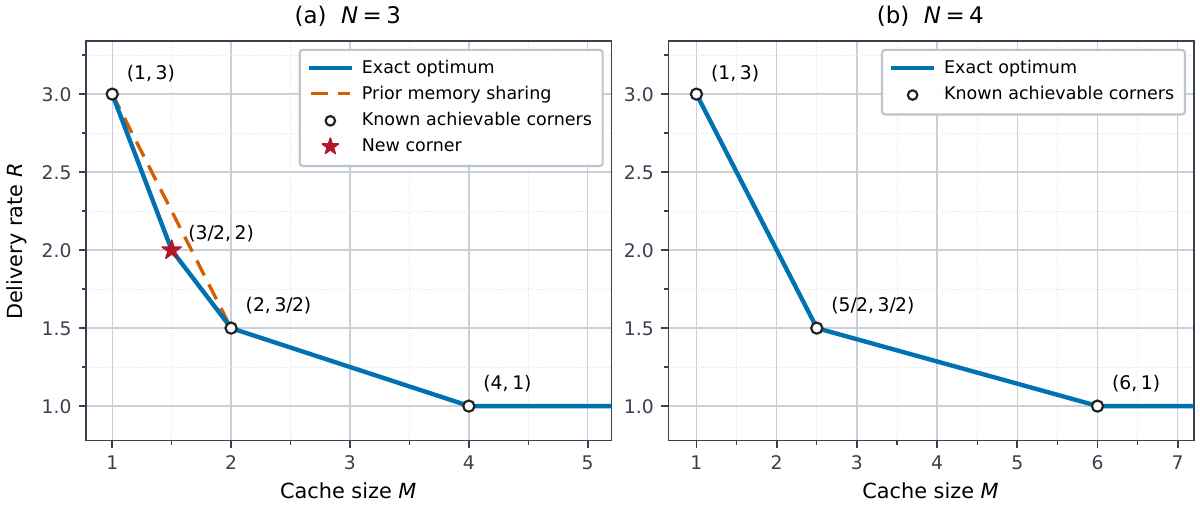}

\caption{Exact memory--rate tradeoffs for three users. (a) For $N=3$,
the new corner $(3/2,2)$ lies below memory sharing between the known
achievable points. The dashed line shows the previous achievable curve on
$1\le M\le2$; for $M\ge2$, it coincides with the exact optimum.
(b) For $N=4$, the converse matches the known
achievable points and their memory-sharing segments. The curves are
given by \cref{thm:main} and continue at $R=1$ for larger memory budgets.}
\label{fig:tradeoff}
\end{figure}

\section{Independence and Recovery Identities}\label{sec:independence}

Every cache is jointly independent of the entire library. For a
nonconstant demand, the broadcast is also independent of the library.
The cache and broadcast together recover the requested file, so they
form two perfect shares of that file.

\subsection{Symmetry and Relabeling}

\begin{proposition}[Symmetrization]\label{prop:sym}
Time sharing over all file permutations and all user permutations
\cite{tian2018symmetry}
preserves the feasible normalized memory--rate budget and yields a code
with the following three properties. Here $\mathcal I\subseteq[N]$ and
$\mathcal K\subseteq[3]$ are file and user index sets.
\begin{enumerate}[label=(\roman*)]
\item Every placement entropy $H(W_{\mathcal I},Z_{\mathcal K})$ is invariant under separate permutations of the file labels and cache labels.

\item An entropy containing $X_{\bm d}$ is invariant under the simultaneous action that applies the same relabeling to its variables and to the demand $\bm d$.
\item If users 1 and 2 request the same file and user 3 requests a different
file, entropy expressions involving
files, caches, and the broadcast for that demand are invariant under
exchanging $Z_1$ and $Z_2$ while keeping the broadcast index fixed.
\end{enumerate}
\end{proposition}

\begin{proof}
Let $\mathcal G$ be the product of the file- and user-permutation groups,
and put $t=|\mathcal G|=6N!$. Split each new file into $t$ independent
subfiles of entropy $F$ and run one independently randomized copy of
the original code for each $(\pi,\sigma)\in\mathcal G$. In that copy,
original file $i$ represents new file $\pi(i)$ and original user $j$
represents new user $\sigma(j)$. A new demand $\bm d$ is served as
$d^{\pi,\sigma}_j=\pi^{-1}(d_{\sigma(j)})$.

Concatenating the outputs gives zero-error recovery. Independence
across copies preserves joint security of the unrequested subfiles.
The product cache and broadcast alphabets have logarithmic sizes at
most $t\max_j\log_2|\mathcal Z_j|$ and
$t\max_{\bm e}\log_2|\mathcal X_{\bm e}|$, respectively, while the new
file entropy is $tF$. Thus the normalized budget is unchanged.
Relabeling files and users permutes the independent copies, proving
the entropy invariances. A relabeling that fixes a demand also fixes
its broadcast index.
\end{proof}

\subsection{Library Independence}

\begin{lemma}[Each Cache Is Independent of the Entire Library]\label{lem:cacheind}
For every user $k$,
\[
  I(W_{[N]};Z_k)=0.
\]
\end{lemma}

\begin{proof}

Fix a user $k$ and a file $W_a$.  Since $N\ge2$, choose a demand under which user $k$ requests some $W_b$ with $b\ne a$.  The joint security condition \eqref{eq:privacy} implies
\[
  I(W_a;Z_k,X_{\bm d})=0,
\]
and hence $I(W_a;Z_k)=0$.

Choose a demand under which the same user requests $W_a$, and set $O=(Z_k,X_{\bm d})$.  Security gives $W_{[N]\setminus\{a\}}\ind O$, and decoding makes $W_a$ a deterministic function of $O$.  Therefore
\begin{align*}
  H(W_{[N]}\mid O)
  &=H(W_{[N]\setminus\{a\}}\mid O)\\
  &=H(W_{[N]\setminus\{a\}})=(N-1)F,
\end{align*}
so $I(W_{[N]};O)=F$.  Decodability and the identity $I(W_a;Z_k)=0$ also give
\begin{align*}
  I(W_{[N]};X_{\bm d}\mid Z_k)
  &\ge I(W_a;X_{\bm d}\mid Z_k)\\
  &=H(W_a\mid Z_k)-H(W_a\mid Z_k,X_{\bm d})\\
  &=F,
\end{align*}
where the last equality uses $I(W_a;Z_k)=0$ and \eqref{eq:decode}.  Applying the chain rule to $O=(Z_k,X_{\bm d})$ yields
\[
  F
  =I(W_{[N]};Z_k)+I(W_{[N]};X_{\bm d}\mid Z_k)
  \ge I(W_{[N]};Z_k)+F.
\]
Thus $I(W_{[N]};Z_k)=0$.
\end{proof}

\begin{lemma}[A Nonconstant-Demand Delivery Is Independent of the Library]\label{lem:xind}
If the demand $\bm d$ contains at least two distinct requested files, then
\[
  I(W_{[N]};X_{\bm d})=0.
\]
\end{lemma}

\begin{proof}
Let user $k$ request $W_a$. Write $V=W_{[N]\setminus\{a\}}$ for its
unrequested file tuple and $O=(Z_k,X_{\bm d})$ for its observation.
Security gives $I(V;O)=0$. Since $W_a$ is a deterministic function of
$O$, the pair $(W_a,X_{\bm d})$ is also a deterministic function of $O$.
The data-processing inequality therefore gives
\[
  I(V;W_a,X_{\bm d})=0.
\]
In particular, $I(V;X_{\bm d}\mid W_a)=0$, and the chain rule implies
\[
  I(W_{[N]};X_{\bm d})=I(W_a;X_{\bm d}).
\]
Because the demand contains at least two distinct requested files, another user $k'$ requests a file $W_b$ with $b\ne a$.  Relative to that user, $W_a$ belongs to the jointly protected tuple of unrequested files.  Thus \eqref{eq:privacy} gives
\[
  I(W_a;Z_{k'},X_{\bm d})=0,
\]
and hence $I(W_a;X_{\bm d})=0$.  The claimed joint library independence follows.
\end{proof}

\begin{proposition}[Labeled Perfect-Share Representation]\label{prop:perfect-share}
Fix a nonconstant demand $\bm d$ and a user $k$.  Then
\[
  I(W_{[N]};Z_k)=I(W_{[N]};X_{\bm d})=0.
\]
The pair $(Z_k,X_{\bm d})$ recovers $W_{d_k}$ and satisfies
\[
  I\bigl(W_{[N]\setminus\{d_k\}};
    W_{d_k},Z_k,X_{\bm d}\bigr)=0.
\]
Thus $Z_k$ and $X_{\bm d}$ are two perfect shares of $W_{d_k}$.
\end{proposition}

\begin{proof}
The individual independence statements follow from
\cref{lem:cacheind,lem:xind}. By decoding, adding $W_{d_k}$ to
$(Z_k,X_{\bm d})$ leaves its information about the unrequested files
unchanged, so the joint security assertion follows from \eqref{eq:privacy}.
\end{proof}

\subsection{Entropy Identities}\label{sec:entropy-identities}

Secret peeling separates the entropy of a recovered secret from the
uncertainty left in the observation after recovery. Same-secret
uncrossing combines several cache--broadcast pairs that recover the
same secret and counts the common broadcast entropy only once.

\begin{lemma}[Secret Peeling]\label{lem:peel}
Let $S,T,U$ be finite-alphabet random variables. If $S\ind T$ and
$H(S\mid T,U)=0$, then
\[
  H(U\mid T)=H(S)+H(U\mid S,T).
\]
\end{lemma}

\begin{proof}
Recoverability and the chain rule give
$H(U\mid T)=H(S,U\mid T)=H(S\mid T)+H(U\mid S,T)$.
Use $S\ind T$ to replace $H(S\mid T)$ by $H(S)$.
\end{proof}

\begin{definition}[Conditional Total Correlation]
Let $r$ be a nonnegative integer and let $U,Z_1,\ldots,Z_r$ be
finite-alphabet random variables. Put $\mathcal R=\{1,\ldots,r\}$.
The difference between the sum of the individual conditional entropies
and their joint conditional entropy is the conditional total correlation:
\[
  \TC(Z_{\mathcal R}\mid U)
  :=\sum_{i\in\mathcal R}H(Z_i\mid U)
    -H(Z_{\mathcal R}\mid U)
  \ge0.
\]
For $r=0$, the tuple and sum are empty, so its total correlation is zero.
For $r=2$, this reduces to $I(Z_1;Z_2\mid U)$.
\end{definition}

\begin{corollary}[Two-Share Form]\label{cor:uncross}
Let $A$ be a file of entropy $F$. If $A$ is recoverable from both
$(Z_1,X)$ and $(Z_2,X)$ and $A\ind X$, then
\begin{align}
  H(Z_1)+H(Z_2)+H(X)
  ={}&2F+H(Z_1,Z_2,X\mid A)\nonumber\\
  &+I(Z_1;X)+I(Z_2;X)\nonumber\\
  &+I(Z_1;Z_2\mid A,X).
  \label{eq:same-secret}
\end{align}
\end{corollary}

\begin{proof}
For $i=1,2$, recovery of $A$ gives
$H(Z_i,X)=F+H(Z_i,X\mid A)$. Also,
\[
\begin{aligned}
&H(Z_1,X\mid A)+H(Z_2,X\mid A)\\
&\quad=H(Z_1,Z_2,X\mid A)+H(X\mid A)
       +I(Z_1;Z_2\mid A,X).
\end{aligned}
\]
Add the two recovery identities and use $H(X\mid A)=H(X)$ and
$H(Z_i)=H(Z_i,X)-H(X)+I(Z_i;X)$ to obtain
\eqref{eq:same-secret}.
\end{proof}

\section{Local Converse Bounds}
\label{sec:local}

\subsection{A Repeated Demand and Its Coalition Coordinates}

We bound the information in cache coalitions using a demand in which two users request the same file.  Choose two distinct library files $A$ and $B$, each of entropy $F$.  The demand label $(A,A,B)$ means that users 1 and 2 request file $A$ and user 3 requests file $B$.  Fix this demand throughout the section and write
\[
 X=X_{(A,A,B)},
 \qquad
 \bZ=(Z_1,Z_2,Z_3).
\]
All identities are evaluated on the symmetric code supplied by \cref{prop:sym}.
For random tuples $U,V$, write $U\rightsquigarrow V$ when $V$ is a deterministic function of $U$.  By \cref{prop:perfect-share}, the three cache--broadcast pairs recover their requested files:
\[
 (Z_1,X)\rightsquigarrow A,
 \qquad
 (Z_2,X)\rightsquigarrow A,
 \qquad
 (Z_3,X)\rightsquigarrow B.
\]

For $N\ge2$, define the two-cache coalition coordinate
\[
 \sigma_B=I(B;Z_1,Z_2\mid A),
\]
and, for $N\ge3$, choose a third distinct file $C$.  In the three-core bound, $Q$ denotes the tuple of the remaining $m=N-3$ files, so $H(Q)=mF$.  The tuple is empty when $N=3$.  Define the three-cache coalition coordinate by
\[
 \sigma_Q=I(Q;\bZ\mid A,B,C).
\]
The distinguished files $A,B,C$ form the three-file core.  Recovery and
security constrain the information that the cache coalitions can hold
about the remaining files.

\begin{lemma}[Minimum Share Entropy]
\label{lem:minshare}
For every nonconstant demand and every user,
\[
 H(Z_k)\ge F,
 \qquad
 H(X_{\bm d})\ge F.
\]
Consequently, $M\ge1$ and $R\ge1$ whenever $N\ge2$.
\end{lemma}

\begin{proof}
For a user $k$ requesting $W_a$, each share $Z_k$ and $X_{\bm d}$ is
independent of $W_a$, and the pair recovers it.  Hence
\[
 F=I(W_a;Z_k\mid X_{\bm d})\le H(Z_k).
\]
Exchanging the shares gives $F\le H(X_{\bm d})$.
\end{proof}

\subsection{Delivery and Cache-Pair Information}

The broadcast entropy is at least $F$ plus the uncertainty about $B$ that remains after the two users requesting $A$ pool their caches and are given $A$.

\begin{lemma}[Delivery Completion]
\label{lem:completion}
For a finite symmetric code with $N\ge3$, choose a third file
$C$ distinct from $A$ and $B$.  Let
\[
 d=H(B\mid A,Z_1,Z_2).
\]
The gap $H(X)-F-d$ admits the exact nonnegative decomposition
\begin{align}
H(X)-F-d
={}&I(Z_3;X)+I(Z_1;X\mid B,Z_3)\nonumber\\
&+I(C;X\mid A,B,Z_1,Z_3)\nonumber\\
&+H(X\mid A,B,C,Z_2,Z_3)\ge0.
\label{eq:completion}
\end{align}
In particular,
\[
 H(X)\ge F+d.
\]
For $N=2$ with library $(A,B)$, the same repeated demand
$X=X_{(A,A,B)}$ and the same residual $d=H(B\mid A,Z_1,Z_2)$ satisfy the exact specialization
\begin{equation}
\begin{aligned}
H(X)-F-d
={}&I(Z_3;X)+I(Z_1;X\mid B,Z_3)\\
&+H(X\mid A,B,Z_2,Z_3)\ge0.
\end{aligned}
\label{eq:completion-two-file}
\end{equation}
Consequently, $H(X)\ge F+d$ also holds for $N=2$.
\end{lemma}

\begin{proof}

Independence $B\ind Z_3$ and decoding from $(Z_3,X)$ give
\[
 H(X)=I(Z_3;X)+F+H(X\mid B,Z_3).
\]
Expanding with respect to $Z_1$ and using recovery of $A$ give
\begin{align*}
H(X\mid B,Z_3)
={}&I(Z_1;X\mid B,Z_3)\\
&+H(A\mid B,Z_1,Z_3)\\
&+H(X\mid A,B,Z_1,Z_3).
\end{align*}
Placement symmetry gives
\begin{align*}
 H(A\mid B,Z_1,Z_3)
 &=H(B\mid A,Z_1,Z_3)\\
 &=H(B\mid A,Z_1,Z_2)=d,
\end{align*}
where the two equalities use the file swap $A\leftrightarrow B$ and the cache swap $2\leftrightarrow3$, respectively.  Expanding with respect to $C$ and then exchanging users $1$ and $2$, which preserves the fixed demand, gives
\begin{align*}
H(X\mid A,B,Z_1,Z_3)
={}&I(C;X\mid A,B,Z_1,Z_3)\\
&+H(X\mid A,B,C,Z_2,Z_3).
\end{align*}
Substitution proves \eqref{eq:completion}, whose four terms are nonnegative.

For $N=2$, stop the expansion before introducing $C$. Exchanging
users 1 and 2 in $H(X\mid A,B,Z_1,Z_3)$ gives
$H(X\mid A,B,Z_2,Z_3)$, proving \eqref{eq:completion-two-file}.
\end{proof}

Because the two users recover the same file $A$, same-secret uncrossing relates their cache entropies and the broadcast entropy to the residual uncertainty about $B$.

\begin{lemma}[Repeated-Request Uncrossing]
\label{lem:resource-uncross}
Let $T=(Z_1,Z_2,X)$ and $d=H(B\mid A,Z_1,Z_2)$.  Define the nonnegative remainder
\begin{align}
\mathsf U_3={}&I(Z_1;X)+I(Z_2;X)
 +I(Z_1;Z_2\mid A,X)\nonumber\\
&+I(B;X\mid A,Z_1,Z_2)\ge0.
\label{eq:U3}
\end{align}
Then the cache and broadcast entropies satisfy the exact identity
\begin{align}
&H(Z_1)+H(Z_2)+H(X)+d\nonumber\\
&\qquad=F+H(A,B,T)+\mathsf U_3.
\label{eq:resource-uncross}
\end{align}
\end{lemma}

\begin{proof}
The two-share form of same-secret uncrossing applied to the users requesting $A$ gives
\begin{align*}
H(Z_1)+H(Z_2)+H(X)
={}&2F+H(T\mid A)\\
&+I(Z_1;X)+I(Z_2;X)\\
&+I(Z_1;Z_2\mid A,X).
\end{align*}
Moreover,
\[
 H(T\mid A)-H(T\mid A,B)
 =F-H(B\mid A,Z_1,Z_2,X),
\]
and
\[
 d-H(B\mid A,Z_1,Z_2,X)
 =I(B;X\mid A,Z_1,Z_2).
\]
Substitute these two identities and use
$H(A,B,T)=2F+H(T\mid A,B)$.
\end{proof}

\subsection{The Entropy of a Repeated-Demand Observation}

Write
\[
 \mathcal Y=(Z_1,Z_2,X),\qquad
 \mathcal L=(Z_1,Z_2,Z_3,X).
\]
The pooled observation $\mathcal Y$ recovers $A$, and $\mathcal L$ also
recovers $B$.  Their overlap and the security of $B$ at $(Z_2,X)$ force
the individual observation to carry at least three file units of entropy.

\begin{lemma}[Repeated-Demand Observation Budget]\label{lem:repeated-view}
For the fixed demand $(A,A,B)$, define the residual entropies
\[
 e=H(X\mid A,Z_1,Z_2),\qquad
 w=H(Z_3\mid B,Z_1,Z_2,X).
\]
Then
\begin{equation}
 H(Z_2,X)\ge3F+e+w.
 \label{eq:repeated-view-budget}
\end{equation}
\end{lemma}

\begin{proof}
Let $b=H(B\mid\mathcal Y)$.  Both $(Z_1,X)$ and $(Z_2,X)$
recover $A$, while $A$ is independent of $X$.  Submodularity after
conditioning on $A$ gives
\begin{equation}
 H(\mathcal Y)\le H(Z_1,X)+H(Z_2,X)-H(X)-F.
 \label{eq:common-file-overlap}
\end{equation}
Next apply submodularity, conditioned on $B$, to $\mathcal Y$ and
$(Z_1,Z_3,X)$.  Their common observation $(Z_1,X)$ is independent of $B$,
and $(Z_1,Z_3,X)$ recovers $B$.  Removing the recovered file gives
\begin{align}
 H(\mathcal L)
 &\le H(\mathcal Y)+b+H(Z_1,Z_3,X)\nonumber\\
 &\quad-H(Z_1,X)-F.
 \label{eq:private-file-overlap}
\end{align}
Finally, $Z_3$ is independent of $B$ and $(Z_3,X)$ recovers $B$, so
\begin{align}
 H(Z_1,Z_3,X)
 &=H(B,Z_1,Z_3)+H(X\mid B,Z_1,Z_3)\nonumber\\
 &\le H(B,Z_1,Z_3)+H(X\mid B,Z_3)\nonumber\\
 &=H(B,Z_1,Z_3)+H(X\mid Z_3)-F\nonumber\\
 &\le H(B,Z_1,Z_3)+H(X)-F.
 \label{eq:third-user-peeling}
\end{align}
Adding these three inequalities cancels $H(Z_1,X)$ and $H(X)$:
\[
 H(\mathcal L)
 \le H(Z_2,X)+H(B,Z_1,Z_3)+b-3F.
\]
Placement symmetry gives $H(B,Z_1,Z_3)=H(A,Z_1,Z_2)$.
The chain rule and the two recovered files give
\[
 H(\mathcal L)=H(A,Z_1,Z_2)+e+b+w.
\]
Substitution proves \eqref{eq:repeated-view-budget}.
\end{proof}

The three file units in \eqref{eq:repeated-view-budget} come from common
recovery of $A$, security of $B$ at the common observation in the second
comparison, and recovery of $B$ from its two shares.  The residuals $e,w$
keep track of the broadcast and cache uncertainty left after the indicated
recoveries.
Joint security and recovery of $A$ also give
$H(Z_2,X\mid W_{[N]})=H(Z_2,X\mid A)$.
Thus the observation budget can be written as
\begin{equation}
 H(Z_2,X\mid W_{[N]})\ge2F+e+w.
 \label{eq:private-view-randomness}
\end{equation}
Even after the library is fixed, user 2's observation retains at least
two file units of entropy from the server's randomness, in addition to
the residuals $e,w$.

A third file $C$ enters through the placement.  Once $A$ is given, the
full cache tuple leaves the same uncertainty about $B$ and $C$.  The
broadcast must complete recovery of $B$.

\begin{lemma}[File-Residual Comparison]\label{lem:file-residual}
For a third file $C$ distinct from $A,B$,
\begin{equation}
 H(X\mid A,Z_1,Z_2)\ge H(C\mid\mathcal L).
 \label{eq:file-residual-comparison}
\end{equation}
\end{lemma}

\begin{proof}
Decoding at user 3 and placement symmetry give
\begin{align*}
 H(X\mid A,Z_1,Z_2)
 &\ge H(X\mid A,\bZ)\\
 &\ge H(B\mid A,\bZ)\\
 &=H(C\mid A,\bZ)\\
 &\ge H(C\mid A,\bZ,X)
 =H(C\mid\mathcal L).
\end{align*}
Placement symmetry justifies the exchange of $B,C$.
The last equality uses recovery of $A$ from $(Z_2,X)$.
\end{proof}

\subsection{The Three-Core Bound}

To retain the information that other caches can supply beyond user 2's
observation, define
\begin{equation}
 K_3=H(Z_1,Z_3\mid B,C,Z_2,X).
 \label{eq:hidden-cache-entropy}
\end{equation}
This hidden cache entropy also bounds the information about files outside
the core.

\begin{lemma}[Three-Core Cycle Bound]
\label{lem:threecore}
For the library $(A,B,C,Q)$, with $Q$ containing $m=N-3$ files, and demand $(A,A,B)$,
\begin{equation}
 H(A,B,Z_1,Z_2,X)+H(Q\mid A,B,C,\bZ)
 \ge(m+5)F.
\label{eq:threecore-joint}
\end{equation}
Consequently,
\begin{equation}
 2H(Z_1)+H(X)\ge5F+\sigma_B+\sigma_Q.
\label{eq:threecore-synergy}
\end{equation}
\end{lemma}

\begin{proof}
Let $r=H(C\mid\mathcal L)$ and $b=H(B\mid\mathcal Y)$, and use $e,w$ from
\cref{lem:repeated-view}.  User 2's joint security gives
$(B,C)\ind(Z_2,X)$.  Since $\mathcal L$ recovers $B$,
\begin{align*}
 K_3&=H(\mathcal L\mid B,C)-H(Z_2,X\mid B,C)\\
    &=H(\mathcal L)-2F+r-H(Z_2,X).
\end{align*}
The tuple $\mathcal Y$ recovers $A$, and $B$ is recovered when $Z_3$ is
added.  Thus
\begin{align*}
 H(A,B,\mathcal Y)&=H(\mathcal Y)+b,\\
 H(\mathcal L)&=H(\mathcal Y)+b+w.
\end{align*}
Subtracting gives the entropy accounting identity
\begin{equation}
 H(A,B,\mathcal Y)-K_3
 =2F+H(Z_2,X)-w-r.
 \label{eq:core-cut-accounting}
\end{equation}
Apply \cref{lem:repeated-view,lem:file-residual}:
\begin{equation}
 H(A,B,\mathcal Y)-K_3
 \ge5F+e-r\ge5F.
 \label{eq:threecore-local-seed}
\end{equation}
The security of $(B,C)$ supplies $2F$, the observation budget supplies
$3F$, and the file-residual comparison covers the remaining uncertainty
about $C$ by $e$.

For the remaining file tuple $Q$, user 2's joint security and decoding
make $Q$ independent of $(A,B,C,Z_2,X)$.  Hence
\begin{align}
 \sigma_Q
 &\le I(Q;\bZ,X\mid A,B,C)\nonumber\\
 &=I(Q;Z_1,Z_3\mid B,C,Z_2,X)
 \le K_3.
 \label{eq:tail-information-budget}
\end{align}
Together with \eqref{eq:threecore-local-seed}, this gives
$H(A,B,\mathcal Y)\ge5F+\sigma_Q$.
Since $H(Q\mid A,B,C,\bZ)=mF-\sigma_Q$, the joint-entropy
form \eqref{eq:threecore-joint} follows.
Finally, $d=F-\sigma_B$ and repeated-request uncrossing give
\[
 2H(Z_1)+H(X)=H(A,B,\mathcal Y)+\sigma_B+\mathsf U_3.
\]
The remainder $\mathsf U_3$ is nonnegative, proving
\eqref{eq:threecore-synergy}.  Appendix~\ref{app:exchange-audit}
retains the exact slacks for the equality analysis.
\end{proof}

\subsection{A Two-Cache Residual Bound}

Once their common requested file is known, two caches can reveal information about the other files.  This joint information is bounded by one cache's entropy beyond a single file.

\begin{lemma}[Two-Cache Residual Bound]
\label{lem:twocache}
For $N\ge2$ and demand $(1,1,2)$, let $V=W_{2:N}$ be the tuple of files other than $W_1$, and write $X=X_{112}$.  Then
\begin{equation}
 I(V;Z_1,Z_2\mid W_1)\le H(Z_2)-F.
\label{eq:twocache-synergy}
\end{equation}
Equivalently,
\begin{equation}
 H(Z_2)+H(V\mid W_1,Z_1,Z_2)\ge NF.
\label{eq:twocache-residual}
\end{equation}
\end{lemma}

\begin{proof}
Combining user-$1$ security with its decoding of $W_1$ gives
\[
 I(V;W_1,Z_1,X)=0.
\]
Hence
\begin{align*}
I(V;Z_1,Z_2\mid W_1)
&\le I(V;Z_1,Z_2,X\mid W_1)\\
&=I(V;Z_2\mid W_1,Z_1,X)\\
&\le H(Z_2\mid W_1,X).
\end{align*}
User $2$ decodes $W_1$ from $(Z_2,X)$, and user-$3$ security gives $I(W_1;X)=0$.  Decoded-secret peeling therefore gives
\[
 H(Z_2\mid X)=F+H(Z_2\mid W_1,X),
\]
and conditioning reduces entropy to give
\[
 H(Z_2\mid W_1,X)\le H(Z_2)-F.
\]
This proves \eqref{eq:twocache-synergy}.  Since
\[
 I(V;Z_1,Z_2\mid W_1)
 =(N-1)F-H(V\mid W_1,Z_1,Z_2),
\]
the residual form \eqref{eq:twocache-residual} follows.
\end{proof}

\subsection{The Three-File Corner}
\label{sec:small}

For three files, delivery completion and the three-core bound give two supporting faces meeting at $(M,R)=(3/2,2)$.  Equality at their intersection constrains the file information held by pairs and triples of caches.  These equality conditions guide the construction in \cref{sec:n3ach}.
Define the conditional information in the cache pair and the remaining uncertainty by
\begin{align*}
 p&=I(W_2;Z_1,Z_2\mid W_1),\\
 d&=H(W_2\mid W_1,Z_1,Z_2)=F-p.
\end{align*}
With $Q$ empty, the three-core bound gives
\begin{equation}
 2H(Z_1)+H(X_{112})\ge5F+p.
\label{eq:n3first}
\end{equation}
Since $p\ge0$, the alphabet bounds yield the supporting face
\[
 2M+R\ge5.
\]

Delivery completion gives
\[
 H(X_{112})\ge F+d=2F-p.
\]
Adding this bound to \eqref{eq:n3first} eliminates $p$:
\[
 2H(Z_1)+2H(X_{112})\ge7F,
\]
which gives the supporting face
\[
 M+R\ge\frac72.
\]
The cancellation comes from splitting one file's entropy into cache-pair information and residual uncertainty:
\[
 \underbrace{I(W_2;Z_1,Z_2\mid W_1)}_{\text{coalition information }p}
 +
 \underbrace{H(W_2\mid W_1,Z_1,Z_2)}_{\text{residual uncertainty }F-p}
 =F.
\]
The two new faces meet at
\[
 2M+R=5,
 \qquad
 M+R=\frac72
 \quad\Longrightarrow\quad
 (M,R)=\left(\frac32,2\right).
\]
The construction in \cref{sec:n3ach} attains this intersection.

\begin{corollary}[Three-File Equality Conditions]
\label{cor:n3fingerprint}
For a finite symmetric three-file code, let $q=I(W_3;\bZ\mid W_1,W_2)$ measure the information that all three caches contain about the third file when the first two files are given.  Define the code's actual normalized alphabet sizes by
$M_{\mathrm a}:=\max_k\log_2|\mathcal Z_k|/F$ and
$R_{\mathrm a}:=\max_{\bm d}\log_2|\mathcal X_{\bm d}|/F$.  On the face
$2M_{\mathrm a}+R_{\mathrm a}=5$, $1\le M_{\mathrm a}\le3/2$,
\[
 H(Z_1)=M_{\mathrm a}F,
 \qquad
 H(X_{112})=R_{\mathrm a}F,
 \qquad
 p=0.
\]
On the face $M_{\mathrm a}+R_{\mathrm a}=7/2$, $3/2\le M_{\mathrm a}\le2$, the same equalities $H(Z_1)=M_{\mathrm a}F$ and $H(X_{112})=R_{\mathrm a}F$ hold, and
\[
 p=\left(M_{\mathrm a}-\frac32\right)F.
\]
At $(M_{\mathrm a},R_{\mathrm a})=(3/2,2)$,
\[
 p=0,
 \qquad
 q=F
\]
and
\[
 H(X_{123})=\log_2|\mathcal X_{123}|=2F.
\]
For symmetric code sequences approaching either face, the corresponding normalized entropy equalities and $p$ formula hold with an $o(1)$ term.  A sequence approaching $(3/2,2)$ also satisfies $q/F\to1$ and $H(X_{123})/F\to2$.

At the corner, file symmetry and $p=0$ imply that, after conditioning on one file, the cache pair has zero mutual information about each of the other files.  The equality $q=F$ means that all three caches, together with two files, determine the remaining file.  The construction in \cref{sec:n3ach} realizes these conditions.
\end{corollary}

\begin{proof}
On $2M_{\mathrm a}+R_{\mathrm a}=5$, the alphabet bounds and \eqref{eq:n3first} give
\[
 5F\ge2H(Z_1)+H(X_{112})\ge5F+p.
\]
Thus $p=0$, and both $Z_1$ and $X_{112}$ attain their alphabet entropy bounds.

On $M_{\mathrm a}+R_{\mathrm a}=7/2$, the sum of \eqref{eq:n3first} and delivery completion gives
\[
 2H(Z_1)+2H(X_{112})\ge7F.
\]
The alphabet bounds give equality.  Consequently, \eqref{eq:n3first} and delivery completion are equalities, and $H(Z_1)=M_{\mathrm a}F$ and $H(X_{112})=R_{\mathrm a}F$.  Using $H(X_{112})=2F-p$ and $R_{\mathrm a}=7/2-M_{\mathrm a}$ yields
\[
 p=\left(M_{\mathrm a}-\frac32\right)F.
\]

At $(3/2,2)$, the face $2M_{\mathrm a}+R_{\mathrm a}=5$ gives $p=0$.  Successively expanding the all-distinct delivery entropy and applying each user's decoding condition gives
\begin{align*}
H(X_{123})
&\ge H(W_1\mid Z_1)
 +H(W_2\mid W_1,Z_1,Z_2)\\
&\quad+H(W_3\mid W_1,W_2,\bZ)\\
&=3F-p-q.
\end{align*}
The alphabet constraints $H(X_{123})\le2F$ and $q\le F$ force $q=F$ and $H(X_{123})=2F$.  Applying the same bounds after division by $F$ gives the limiting equalities for code sequences.
\end{proof}

\section{A Three-File Construction at \texorpdfstring{$(3/2,2)$}{(3/2,2)}}\label{sec:n3ach}

\subsection{Equality Conditions and Parameters}

We construct a code that attains the intersection of the two three-file
supporting faces.  The equality conditions in \cref{cor:n3fingerprint}
guide the placement: a finite symmetric code at this point must satisfy
\begin{align*}
 p&=I(W_2;Z_1,Z_2\mid W_1)=0,\\
 q&=I(W_3;Z_1,Z_2,Z_3\mid W_1,W_2)=F.
\end{align*}
Thus, given one file, a cache pair has no information about either unknown
file individually. Given two files, the three caches determine the remaining
file. Our placement also lets the three caches recover both remaining files
when any one file is given.
Each file consists of two field symbols, each cache stores three symbols,
and each nonconstant-demand broadcast contains four symbols. The delivery
rule leaves two independent combinations of the requested file after key
cancellation. Five independent keys mask the complete unrequested tuple.

All arithmetic in the construction is over $\mathbb F_3$.  Take each file
$W_i\in\mathcal V_i=\mathbb F_3^2$ to be a uniform random column, independently
of the other files.  Put
$\mathcal V=\mathcal V_1\oplus\mathcal V_2\oplus\mathcal V_3$, and write
\[
 W=(W_1^{\mathsf T},W_2^{\mathsf T},W_3^{\mathsf T})^{\mathsf T}
 \in\mathbb F_3^6.
\]
Let
\[
 S=(s_{12},s_{13},t_1,s_{23},t_2)^{\mathsf T}\in\mathbb F_3^5
\]
consist of five mutually independent uniform keys, independent of the
library.

\subsection{Block-MDS Placement}

The placement uses four scalar linear combinations of the files, called
syndromes.  Write $12,13,23,123$ for the subsets
$\{1,2\},\{1,3\},\{2,3\},\{1,2,3\}$, respectively.
For each of these labels $J$, define the coefficient row $r_J$ by the
following table and set $g_J=r_JW$:
\[
\begin{array}{c|c}
J&r_J\\ \hline
12&(0,2\,|\,2,0\,|\,2,1)\\
13&(2,0\,|\,1,1\,|\,1,1)\\
23&(2,1\,|\,0,1\,|\,1,0)\\
123&(0,2\,|\,2,1\,|\,0,2).
\end{array}
\]
The caches store the following key-masked combinations:
\begin{equation}
\begin{aligned}
Z_1&=(s_{12},\,s_{13},\,t_1),\\
Z_2&=(s_{12}+g_{12},\,s_{23},\,t_2),\\
Z_3&=(s_{13}+g_{13},\,s_{23}+g_{23},\,t_1+t_2+g_{123}).
\end{aligned}
\label{eq:n3placement}
\end{equation}
A file has two field symbols and each cache has three, giving $M=3/2$.
In each cache, the key-coefficient matrix has rank three.  Conditional on
the library, the cache is therefore uniform on its three-symbol alphabet,
so it is independent of the library.

To describe key cancellation, let $E=\mathbb F_3^5$ be the space of key
coefficient rows, with coordinate basis $(e_1,\ldots,e_5)$.  The rows used by
the three caches span
\begin{align}
C_1&=\langle e_1,e_2,e_3\rangle,\nonumber\\
C_2&=\langle e_1,e_4,e_5\rangle,\nonumber\\
C_3&=\langle e_2,e_4,e_3+e_5\rangle.
\label{eq:keyplanes}
\end{align}
The intersections $C_1\cap C_2$, $C_1\cap C_3$, and $C_2\cap C_3$ are
$\langle e_1\rangle$, $\langle e_2\rangle$, and $\langle e_4\rangle$,
respectively.  The triple intersection is zero, and any two cache subspaces
span $E$.  Write $Z_{k,r}$ for coordinate $r$ of cache $Z_k$.
Cancelling shared key terms between caches recovers the four syndromes:
\[
\begin{aligned}
 Z_{2,1}-Z_{1,1}&=g_{12},\\
 Z_{3,1}-Z_{1,2}&=g_{13},\\
 Z_{3,2}-Z_{2,2}&=g_{23},\\
 Z_{3,3}-Z_{1,3}-Z_{2,3}&=g_{123}.
\end{aligned}
\]
The subscripts $12,13,23,123$ identify the caches involved in each cancellation.
Stack the coefficient rows of these syndromes as
\[
 \Theta=\begin{bmatrix}r_{12}\\r_{13}\\r_{23}\\r_{123}\end{bmatrix}.
\]
The matrix $\Theta$ has rank four, and every $4\times4$ submatrix obtained
by retaining any pair of two-column file blocks is nonsingular.  Equivalently,
after invertible row operations and a separate change of basis inside each
file, it becomes
\begin{equation}
 \Theta_{\rm can}=
 \begin{bmatrix}I_2&0&-I_2\\0&I_2&-I_2\end{bmatrix}.
\label{eq:blockmds}
\end{equation}
The nonsingularity of every two-block submatrix is the maximum-distance-separable
(MDS) property needed here: the four syndromes determine any two file blocks
once the third is known.  In coding terminology, $\Theta_{\rm can}$ is a
$[3,1,3]$ array-MDS parity check.  In suitable file coordinates,
$\Theta_{\rm can}$ maps $(Y_1,Y_2,Y_3)$ to the pair of differences
$(Y_1-Y_3,Y_2-Y_3)$.  Its kernel is the diagonal subspace
$\{(u,u,u):u\in\mathbb F_3^2\}$.

Each cache pair cancels one shared key direction and reveals a scalar
syndrome with nonzero components in all three file blocks.
After conditioning on any one file,
one of the remaining file blocks uniformly masks the contribution of the
other.  In particular,
$I(W_2;Z_1,Z_2\mid W_1)=p=0$.  The three caches reveal all four syndromes
in $\Theta W$, which determine the other two file blocks once any one file
is given.  Thus $I(W_3;Z_1,Z_2,Z_3\mid W_1,W_2)=q=F$.

\subsection{A Decoding Example}

For the all-distinct demand $\bm d=(1,2,3)$, write
\begin{align*}
 W_1&=(a,b)^{\mathsf T},&W_2&=(c,d)^{\mathsf T},\\
 W_3&=(e,f)^{\mathsf T}.
\end{align*}
Transmit the following four symbols:
\[
\begin{aligned}
X_1&=a+s_{12}+s_{13},\\
X_2&=2b+t_1,\\
X_3&=b+c+2d+e+2f+2s_{12}+s_{23},\\
X_4&=2c+d+t_2.
\end{aligned}
\]
User 1 cancels its three cache keys to obtain
\begin{align*}
 a&=X_1-Z_{1,1}-Z_{1,2},\\
 2b&=X_2-Z_{1,3}.
\end{align*}
User 2 obtains
\begin{align*}
 2d&=X_3-(2Z_{2,1}+Z_{2,2}),\\
 2c+d&=X_4-Z_{2,3}.
\end{align*}
User 3 obtains
\[
\begin{aligned}
 e+2f&=Z_{3,1}+Z_{3,2}+2X_1+2X_3,\\
 2f&=Z_{3,3}+2X_2+2X_4.
\end{aligned}
\]
For each user, the two displayed combinations form an invertible system
for its requested file.

\subsection{Cancellation Subspaces and Support Selection}

The general delivery rule must leave each user two independent file
combinations after key cancellation.  Fix a nonconstant demand
$\bm d=(d_1,d_2,d_3)\in[3]^3$.  Its requested-file set is
$J=\{d_1,d_2,d_3\}=\operatorname{supp}(\bm d)$, so $|J|$ is two or three.
All delivery coefficients below are fixed functions of this demand.
For each requested-file set $J$, choose a linear functional $\ell_J$ on $E$
from the table below.  Its listed row acts on $h\in E$ as
$\ell_Jh^{\mathsf T}$.  We choose the broadcast key rows to form a basis
of the hyperplane $H_J=\ker\ell_J\subset E$:
\begin{equation}
\begin{array}{c|c}
J&\ell_J\\ \hline
12&(1,1,2,2,2)\\
13&(1,2,1,2,1)\\
23&(1,1,1,1,2)\\
123&(1,2,0,1,0).
\end{array}
\label{eq:support-hyperplanes}
\end{equation}
For user $k$, the key directions cancellable between its cache and this
broadcast are precisely
$D_k^J=C_k\cap H_J$.  Direct calculation gives
\begin{align}
\dim D_k^J&=2,\nonumber\\
D_i^J\cap D_j^J&=0,\nonumber\\
D_i^J\oplus D_j^J&=H_J\qquad(i\ne j).
\label{eq:partial-spread}
\end{align}
Thus each user can form two independent combinations in which the cache
and broadcast keys cancel.  The cancellation subspaces of any two users
give complementary coordinates on $H_J$.

To track the file terms that remain when keys cancel, let $\mathcal V^*$
denote the space of linear functionals on the file-column space $\mathcal V$.
Its elements are file coefficient rows.  The placement defines a linear map
$\varphi_k:C_k\to\mathcal V^*$: the cache combination with key coefficient
row $h$ has file coefficient row $\varphi_k(h)$.  In the cache bases
of \eqref{eq:keyplanes}, placement \eqref{eq:n3placement} gives
$\varphi_2(e_1)=r_{12}$ and
\begin{align*}
 \varphi_3(e_2)&=r_{13},&\varphi_3(e_4)&=r_{23},\\
 \varphi_3(e_3+e_5)&=r_{123}.
\end{align*}
The remaining values are
$\varphi_1(e_1)=\varphi_1(e_2)=\varphi_1(e_3)=0$ and
$\varphi_2(e_4)=\varphi_2(e_5)=0$.
Since $H_J=D_1^J\oplus D_2^J$, let
$\pi_i:D_3^J\to D_i^J$ be the restriction to $D_3^J$ of the projection
onto summand $D_i^J$, for $i=1,2$.  Every $h\in D_3^J$ then decomposes uniquely as
\[
 h=\pi_1h+\pi_2h,
 \qquad \pi_i h\in D_i^J.
\]
The restrictions of $\varphi_1$ and $\varphi_2$ to $D_1^J$ and $D_2^J$
extend uniquely to a map on $H_J$.  To determine the file terms remaining
at user~3, we compare $\varphi_3$ with that extension on $D_3^J$.
Their difference is the \emph{extension defect}
$\kappa_J:D_3^J\to\mathcal V^*$:
\begin{equation}
 \kappa_J(h)=\varphi_3(h)-\varphi_1(\pi_1h)-\varphi_2(\pi_2h).
\label{eq:kappa}
\end{equation}
For $j\in[3]$, let $\operatorname{pr}_j:\mathcal V^*\to\mathcal V_j^*$ be
projection onto the file-$j$ coefficient block and write
$\kappa_{J,j}=\operatorname{pr}_j\circ\kappa_J$.
The hyperplanes in \eqref{eq:support-hyperplanes} give
\begin{equation}
\begin{aligned}
\kappa_{J,j}=0
 &\quad\Longleftrightarrow\quad j\notin J,\\
\kappa_{J,j}\text{ is an isomorphism}
 &\quad\Longleftrightarrow\quad j\in J.
\end{aligned}
\label{eq:kappa-selector}
\end{equation}
Appendix~\ref{app:n3-closed-form} gives the zero-block equations and rank
calculations.  Among the hyperplanes
satisfying \eqref{eq:partial-spread}, the zero-block and rank conditions
determine the normals for the three two-file requested sets up to
nonzero scaling.

\subsection{Delivery by Linear Interpolation}

The direct sum $H_J=D_1^J\oplus D_2^J$ lets us prescribe the residual file
combinations for users 1 and 2 independently.  We choose two independent
combinations of each user's requested file.  User~3's residual is the
extension defect $\kappa_J$ plus these prescribed combinations.  To cancel
every block outside user~3's request while retaining an invertible requested
block, choose the nonzero scalars
$\alpha_1,\alpha_2\in\mathbb F_3$ as follows:
\begin{equation}
\alpha_i=
\begin{cases}
1,&d_1=d_2\ne d_3,\\
1,&d_i=d_3,\\
2=-1,&\text{otherwise},
\end{cases}
\qquad i\in\{1,2\}.
\label{eq:alpha-rule}
\end{equation}
Let $\iota_j:\mathcal V_j^*\hookrightarrow\mathcal V^*$ embed a coefficient
row into file block $j$, with zero entries in the other blocks.  The projections
$\pi_i$ are isomorphisms by
\eqref{eq:partial-spread}, so the desired residual maps
$\theta_i:D_i^J\to\mathcal V^*$ for users 1 and 2 are
\[
 \theta_i=\alpha_i\,\iota_{d_i}\circ
 \kappa_{J,d_i}\circ\pi_i^{-1},
 \quad i=1,2.
\]
Because $H_J=D_1^J\oplus D_2^J$, there is a unique broadcast file-coefficient
map $b_{\bm d}:H_J\to\mathcal V^*$ with the following values for
$x\in D_1^J$ and $y\in D_2^J$:
\begin{align}
b_{\bm d}(x+y)
 &=\varphi_1(x)+\varphi_2(y)\nonumber\\
 &\quad-\theta_1(x)-\theta_2(y).
\label{eq:b-interpolation}
\end{align}
For users 1 and 2, the residual file maps are
$\varphi_i-b_{\bm d}=\theta_i$ on $D_i^J$, so no unrequested-file
coefficients remain.

For $\ell_J=(1,\ell_2,\ldots,\ell_5)$, take the canonical basis
\[
 v_r=e_{r+1}-\ell_{r+1}e_1,
 \qquad r=1,\ldots,4,
\]
of $H_J$, and transmit
\begin{equation}
 X_{\bm d}
 =\bigl(b_{\bm d}(v_r)W+v_rS:\ r=1,\ldots,4\bigr).
\label{eq:closed-delivery}
\end{equation}
Let $B_{\bm d}\in\mathbb F_3^{4\times6}$ and
$V_{\bm d}\in\mathbb F_3^{4\times5}$ have row $r$ equal to
$b_{\bm d}(v_r)$ and $v_r$, respectively.  The same broadcast is
\[
 X_{\bm d}=B_{\bm d}W+V_{\bm d}S.
\]
For $\bm d=(1,2,3)$, the normal is $\ell_{123}=(1,2,0,1,0)$.
Substituting the corresponding maps into this rule gives exactly the
four broadcast symbols in the decoding example.
Appendix~\ref{app:n3-closed-form} derives the coordinate form of
\eqref{eq:b-interpolation}--\eqref{eq:closed-delivery}.

\subsection{Decoding and Security}

Conditional on the files, uniform independent keys make a linear
observation uniform on a coset of the key image. Decoding and security
follow by tracking the file contributions to this coset, as in linear
secret sharing and coset coding \cite{cramer2000,ozarow1984}.

\begin{lemma}[Quotient-Coset Criterion]\label{lem:coset}
Let $W\in\mathcal V=\bigoplus_j\mathcal V_j$ have independent file blocks
over a finite field.  Let $S$ be uniform on a key space over the same field
and independent of $W$.  Fix linear maps $G$ and $K$ from the file and key
spaces, respectively, into a coordinate space $\mathcal O$.
Consider the linear observation $O=GW+KS$.
Conditional on $W$, the observation is uniform on the
affine coset $GW+\operatorname{im}K$.  Zero-error recovery of the requested
block and joint perfect security of the complete tuple of unrequested blocks
hold if the induced map
\[
 \overline G:\mathcal V\longrightarrow
 \mathcal O/\operatorname{im}K
\]
vanishes on every unrequested block and is injective on the requested
block.
\end{lemma}

\begin{proof}
Uniformity and independence of $S$ make $KS$, conditional on $W$, uniform on
$\operatorname{im}K$, proving the coset statement.  The quotient class of
$O$ is $\overline G(W)$.  Since $\overline G$ vanishes on the unrequested
blocks and is injective on the requested block, the quotient class
determines the requested block.

Changing the unrequested tuple changes $GW$ only by a vector in
$\operatorname{im}K$.  For each fixed requested block, the conditional
distribution of $O$ is therefore unchanged.  Averaging this distribution
over the requested block, which is independent of the complete unrequested
tuple, shows that $O$ is independent of the tuple.
When the quotient space and requested-file space have the same dimension,
injectivity is equivalent to an isomorphism onto the quotient.
\end{proof}

\begin{theorem}[Achievability of the Three-File Corner]\label{thm:n3closed}
The placement \eqref{eq:n3placement} and delivery
\eqref{eq:closed-delivery} satisfy zero-error decoding and perfect file
security for every nonconstant demand.  For a constant demand, directly
transmit the two requested file symbols.  Hence
$(M,R)=(3/2,2)$ is achievable.
\end{theorem}

\begin{proof}
Fix a nonconstant demand $\bm d$ with requested-file set $J$, and fix user
$k\in[3]$.  Let $\mathbf P_k\in\mathbb F_3^{3\times6}$ and
$\mathbf C_k\in\mathbb F_3^{3\times5}$ be the file and key coefficient
matrices of cache $Z_k$.  The cache and broadcast are
\[
 Z_k=\mathbf P_kW+\mathbf C_kS,
 \qquad
 X_{\bm d}=B_{\bm d}W+V_{\bm d}S.
\]
The row spaces of $\mathbf C_k$ and $V_{\bm d}$ are $C_k$ and $H_J$,
respectively.  To apply \cref{lem:coset}, stack the file and key matrices
separately in the observation space
$\mathcal O=\mathbb F_3^7$:
\begin{equation}
 G_{k,\bm d}=
 \begin{bmatrix}\mathbf P_k\\B_{\bm d}\end{bmatrix},
 \qquad
 K_{k,\bm d}=
 \begin{bmatrix}\mathbf C_k\\V_{\bm d}\end{bmatrix}.
\label{eq:GK-stack}
\end{equation}
The stacked observation is
$(Z_k,X_{\bm d})=G_{k,\bm d}W+K_{k,\bm d}S$.  Since
$C_k+H_J=E$, the key matrix has rank five.  Hence the quotient observation
space has dimension two.  Its dual is the annihilator of the key image,
namely the two-dimensional space
\[
 \mathcal A_{k,\bm d}
 =\{\lambda\in\mathcal O^*: \lambda K_{k,\bm d}=0\}.
\]
These are exactly the coefficient rows of key-free observation combinations.
For $h\in D_k^J=C_k\cap H_J$, there are unique row
coefficient vectors $a\in\mathbb F_3^{1\times3}$ and
$c\in\mathbb F_3^{1\times4}$ such that
\begin{equation}
 a\mathbf C_k=h,
 \qquad
 cV_{\bm d}=-h.
\label{eq:ac-h}
\end{equation}
The bases of $C_k$ and $H_J$ make $h\mapsto(a,c)$ an isomorphism
from $D_k^J$ to $\mathcal A_{k,\bm d}$.  An annihilator row acts on
the quotient by
\begin{equation}
\begin{aligned}
 \lambda&\longmapsto\bigl([o]\mapsto\lambda o\bigr),\\
 \mathcal A_{k,\bm d}&\simeq
 (\mathcal O/\operatorname{im}K_{k,\bm d})^*.
\end{aligned}
\label{eq:annihilator-quotient}
\end{equation}
The condition $\lambda K_{k,\bm d}=0$ makes this action independent
of the representative $o$.

The placement gives $a\mathbf P_k=\varphi_k(h)$.  Linearity of
$b_{\bm d}$ and its evaluation on the rows of $V_{\bm d}$ give
\[
 cB_{\bm d}=b_{\bm d}(cV_{\bm d})=-b_{\bm d}(h).
\]
Therefore the annihilator row $(a,c)$ acts on the file-coefficient matrix as
\begin{equation}
\begin{aligned}
 (a,c)G_{k,\bm d}
 &=a\mathbf P_k+cB_{\bm d}\\
 &=\varphi_k(h)-b_{\bm d}(h).
\end{aligned}
\label{eq:quotient-dual-action}
\end{equation}
Under \eqref{eq:annihilator-quotient}, the residual map in
\eqref{eq:quotient-dual-action} is precisely the dual action
of the quotient observation map in \cref{lem:coset}.

For users 1 and 2, restricting \eqref{eq:b-interpolation} to
$D_1^J$ or $D_2^J$ makes \eqref{eq:quotient-dual-action} equal to
$\theta_1$ or $\theta_2$.  For $i=1,2$, the map $\theta_i$ has zero
output in every unrequested file block, while its projection
$\operatorname{pr}_{d_i}\circ\theta_i:D_i^J\to\mathcal V_{d_i}^*$
onto the requested block is an isomorphism.  This follows from
\eqref{eq:kappa-selector}, the invertibility of $\pi_i$, and the
nonzero coefficient $\alpha_i$.
For user~3 and $h\in D_3^J$, substitution into \eqref{eq:kappa} gives
\begin{align*}
 (\varphi_3-b_{\bm d})(h)
 &=\kappa_J(h)
 +\alpha_1\iota_{d_1}\kappa_{J,d_1}(h)\\
 &\quad+\alpha_2\iota_{d_2}\kappa_{J,d_2}(h).
\end{align*}
The file-$j$ block of this residual map is $\kappa_{J,j}$ multiplied by
\[
 1+\sum_{i\le2:\,d_i=j}\alpha_i.
\]
Every block for a file outside $J$ vanishes by \eqref{eq:kappa-selector}.
For each requested file different from $d_3$, \eqref{eq:alpha-rule} makes the
coefficient zero through $1+1+1=0$ or $1+2=0$ in $\mathbb F_3$.  The
requested output block has coefficient one or two and is an isomorphism by
\eqref{eq:kappa-selector}.  For every user, the residual map thus has
zero projection onto each unrequested dual file space and an isomorphism
onto the requested one.  By \eqref{eq:quotient-dual-action}, these projections
are the duals of the restrictions of the quotient observation map.
Finite-dimensional duality therefore shows that the quotient map vanishes
on every unrequested file space and is an isomorphism on the requested
file space.  The quotient-coset criterion proves zero-error decoding and
joint perfect security of the unrequested tuple.

For a constant demand $(a,a,a)$, transmit $X=W_a$.  Let
$W_{-a}=W_{[3]\setminus\{a\}}$.  Since every cache is independent of the
whole library and the files are mutually independent,
\[
\begin{aligned}
I(W_{-a};Z_k,W_a)
 &=I(W_{-a};W_a)+I(W_{-a};Z_k\mid W_a)\\
 &=0.
\end{aligned}
\]
Thus constant-demand security also holds.  A cache uses three symbols per
two-symbol file, and a nonconstant broadcast uses four, giving
$M=3/2$ and worst-case rate $R=2$.
\end{proof}

\section{The General Low-Memory Bound}\label{sec:low}

For every $N\ge4$, the low-memory converse establishes
\[
 3M+(N-1)R\ge3N.
\]
The cache placement supports delivery under both repeated and distinct
demands.  The two demand types constrain the same coalition information in
different ways.  Combining those constraints eliminates the coalition
information and bounds the broadcast-entropy saving by the cache entropy.

Apply \cref{prop:sym} and fix $N\ge4$ together with a finite symmetric code. Put
\[
 z=H(Z_1),\qquad x=H(X_{112}),\qquad y=H(X_{123}),
\]
and define the information about an additional file revealed by two and
three caches, respectively:
\begin{align}
p&=I(W_2;Z_1,Z_2\mid W_1),\nonumber\\
q&=I(W_3;\bZ\mid W_1,W_2).
\label{eq:pq}
\end{align}
Equivalently,
\begin{align*}
H(W_2\mid W_1,Z_1,Z_2)&=F-p,\\
H(W_3\mid W_1,W_2,\bZ)&=F-q.
\end{align*}
Both coordinates satisfy $0\le p,q\le F$.

\subsection{Information in the Remaining Files}

To extend the core bounds to the whole library, we sum the conditional
entropies of the remaining files. File symmetry makes the marginal
entropies equal. Their sum equals the joint conditional entropy plus the
conditional total correlation.

\begin{lemma}[Exact Symmetric-Tail Aggregation]\label{lem:tail}
Let $t$ be an integer with $0\le t\le N-3$.  An empty tuple has zero
entropy and zero total correlation.  Define the nonnegative correlation term
\begin{align*}
\Gamma_t={}&I(W_{4:3+t};W_3\mid W_1,W_2,\bZ)\\
&+\TC(W_{4:3+t}\mid W_1,W_2,\bZ)\ge0.
\end{align*}
Then
\begin{align}
H(W_{2:N}\mid W_1,Z_{1:2})
={}&(N-1)(F-p)\nonumber\\
&-\TC(W_{2:N}\mid W_1,Z_{1:2}),
\label{eq:tail-two}\\
H(W_{4:3+t}\mid W_{1:3},\bZ)
={}&t(F-q)-\Gamma_t.
\label{eq:tail-three}
\end{align}
\end{lemma}

\begin{proof}
For \eqref{eq:tail-two}, file symmetry makes every conditional marginal
$H(W_s\mid W_1,Z_1,Z_2)$, $s\ge2$, equal to $F-p$.  Subtracting the
conditional total correlation from the sum of these marginals gives the joint
conditional entropy.

For \eqref{eq:tail-three}, let $T$ denote the random file tuple $W_{4:3+t}$.
Symmetry yields the exact marginal-sum identity
\begin{equation}
 t(F-q)=H(T\mid W_1,W_2,\bZ)
 +\TC(T\mid W_1,W_2,\bZ).
\label{eq:tail-marginal-sum}
\end{equation}
Conditioning additionally on $W_3$ gives
\begin{align*}
H(T\mid W_1,W_2,\bZ)
={}&I(T;W_3\mid W_1,W_2,\bZ)\\
&+H(T\mid W_1,W_2,W_3,\bZ).
\end{align*}
Substitution into \eqref{eq:tail-marginal-sum} proves
\eqref{eq:tail-three}.  The correlation slack has the expansion
\[
 \Gamma_t=
 \sum_{s=4}^{3+t}
 I(W_s;W_3,W_{4:s-1}\mid W_1,W_2,\bZ),
\]
a sum of nonnegative conditional mutual informations.

\end{proof}

\subsection{Placement and Delivery Inequalities}

The tail identities turn the two-cache residual bound and the three-core
bound into constraints on the information available to cache pairs and
triples across the whole library.

\begin{lemma}[Bounds on Placement Information]\label{lem:synergy-cone}
For every $N\ge4$,
\begin{align}
z&\ge F+(N-1)p,
\label{eq:budget-cache}\\
2z+x&\ge5F+p+(N-3)q.
\label{eq:budget-three}
\end{align}
The strengthened right-hand sides add
$\TC(W_{2:N}\mid W_1,Z_1,Z_2)$ and $\Gamma_{N-3}$,
respectively. Both correction terms are nonnegative by \cref{lem:tail}.
\end{lemma}

\begin{proof}

Equations~\eqref{eq:twocache-synergy} and \eqref{eq:tail-two} give
\begin{align*}
z&\ge F+I(W_{2:N};Z_1,Z_2\mid W_1)\\
 &=F+(N-1)p\\
 &\quad+\TC(W_{2:N}\mid W_1,Z_1,Z_2),
\end{align*}
which proves \eqref{eq:budget-cache}.  In \eqref{eq:threecore-synergy}, set
$(A,B,C,Q)=(W_1,W_2,W_3,W_{4:N})$.  The two coalition-information terms become
$p$ and
\[
 I(W_{4:N};\bZ\mid W_{1:3})
 =(N-3)q+\Gamma_{N-3},
\]
so the inequality gives \eqref{eq:budget-three}.
\end{proof}

Under distinct requests, successive decoding bounds the broadcast entropy
from below by a sum of conditional file entropies. Each term conditions on
the earlier files and the cache tuple up to the current user.

\begin{lemma}[Successive All-Distinct Peeling]\label{lem:distinct}
Fix an integer $r$ with $1\le r\le3$.  Let users $1,\ldots,r$ request
$W_1,\ldots,W_r$, respectively.  Each remaining user requests one of these
$r$ files.  Let $Y$ denote the random broadcast for this fixed demand.
Then
\begin{align}
H(Y)={}&\sum_{j=1}^{r}
 H(W_j\mid W_{[j-1]},Z_{[j]})\nonumber\\
&+\sum_{j=1}^{r}
 I(Y;Z_j\mid W_{[j-1]},Z_{[j-1]})\nonumber\\
&+H(Y\mid W_{[r]},Z_{[r]}).
\label{eq:successive-peel}
\end{align}
In particular, for the all-distinct demand $(1,2,3)$,
\begin{equation}
 y\ge3F-p-q.
\label{eq:distinct}
\end{equation}
\end{lemma}

\begin{proof}

For $j=1,\ldots,r$, define the random tuple
$C_{j-1}=(W_{[j-1]},Z_{[j-1]})$.  Apply the chain rule after conditioning
additionally on $Z_j$.  Since $(Z_j,Y)$ decodes $W_j$,
\begin{align*}
H(Y\mid C_{j-1})
={}&I(Y;Z_j\mid C_{j-1})
 +H(W_j\mid C_{j-1},Z_j)\\
&+H(Y\mid C_{j-1},Z_j,W_j).
\end{align*}
Iterating for $j=1,\ldots,r$ proves \eqref{eq:successive-peel}.  For $r=3$,
the sum of conditional file entropies is
\[
 F+(F-p)+(F-q),
\]
by cache--library independence and \eqref{eq:pq}.  The other terms in
\eqref{eq:successive-peel} are nonnegative, which gives
\eqref{eq:distinct}.
\end{proof}

\subsection{How Placement Limits Delivery Savings}
\label{sec:low-savings}

We measure cache entropy above the one-file minimum and broadcast entropy
relative to serving the three users separately:
\begin{equation}
 a=z-F,\qquad s=3F-x,\qquad t=3F-y.
\label{eq:cache-and-delivery-savings}
\end{equation}
Here $a\ge0$ by \cref{lem:minshare}. The benchmark $3F$ corresponds to
serving the three users separately with independent one-time pads.
The signed quantities $s$ and $t$ measure the entropy savings for repeated
and all-distinct requests, respectively.
The cache bound \eqref{eq:budget-cache}, the three-core bound
\eqref{eq:budget-three}, and distinct-demand peeling \eqref{eq:distinct}
become
\begin{align}
 (N-1)p&\le a,\label{eq:savings-cache}\\
 s+p+(N-3)q&\le2a,\label{eq:savings-repeated}\\
 t&\le p+q.\label{eq:savings-distinct}
\end{align}
The cache constraint charges $(N-1)p$ bits for the information shared
across the $N-1$ alternative files.  The three-cache information $q$
increases the saving allowed for distinct demands but reduces the saving
allowed by the repeated-demand bound.  These opposing effects link the
two demand types.

The coefficient $N-3$ counts the files outside the three-file core.
Multiplying the distinct-demand inequality by this count cancels the
three-cache information charged in the repeated-demand inequality:
\begin{align}
 s+(N-3)t
 &\le 2a-p-(N-3)q+(N-3)(p+q)\nonumber\\
 &=2a+(N-4)p\nonumber\\
 &\le\left(2+\frac{N-4}{N-1}\right)a
 =\frac{3(N-2)}{N-1}a.
\label{eq:savings-combination}
\end{align}
The last inequality uses $N\ge4$ and \eqref{eq:savings-cache}.

\begin{proof}[Proof of the low-memory supporting line]
For a finite code with entropy bounds $z\le MF$ and $x,y\le RF$, the
definitions give
\[
 a\le(M-1)F,\qquad s,t\ge(3-R)F.
\]
There are $N-2$ delivery weights on the left of
\eqref{eq:savings-combination}, so
\[
 (N-2)(3-R)F
 \le\frac{3(N-2)}{N-1}(M-1)F.
\]
Dividing by $(N-2)F>0$ and rearranging gives
\[
 3M+(N-1)R\ge3N.
\]
For an achievable pair, apply the finite-code inequality with budgets
$M+\epsilon$ and $R+\epsilon$ and let $\epsilon\downarrow0$.
\end{proof}

The same constraints also restrict placement on the optimal face.
For a finite symmetric code with $N>4$, equality forces
$p=(M_{\mathrm a}-1)F/(N-1)$ and $q=2p$, where
$M_{\mathrm a}=\max_k\log_2|\mathcal Z_k|/F$ is the actual normalized
cache alphabet size. At $M_{\mathrm a}=(N+1)/2$, these values are
$p=F/2$ and $q=F$. The three-file corner instead has $p=0,q=F$:
its placement provides the same third-file information through three
caches while keeping the pair coordinate zero. Appendix~\ref{app:exchange-audit}
proves the full equality conditions, including the four-file case.

\section{High-Memory Converse}
\label{sec:high}

For every $N\ge3$, broadcasts for several repeated demands give
$M+(3N-5)R\ge5N-7$.  With one common cache and the file requested by the
other two users, these broadcasts determine the whole library.  Security
then bounds the entropy needed for this recovery.  Two overlapping
cache--broadcast collections yield the bound while a shared conditional
entropy cancels.

\subsection{Recovery Stars and Delivery Excess}

Fix $N\ge3$ and a finite code.  To recover several files with the same user's cache,
we collect broadcasts in which that user changes its request while the
other two request $W_1$.  Define the random broadcasts $A_i$ for
$i=2,\ldots,N$, $B_i$ for $i=3,\ldots,N$, and $C$ by
\[
 A_i\triangleq X_{(1,1,i)},
 \qquad
 B_i\triangleq X_{(1,i,1)},
 \qquad
 C\triangleq X_{(2,1,1)}.
\]
With cache $Z_3$, the broadcasts $A_2,\ldots,A_N$ recover $W_2,\ldots,W_N$,
respectively.  These cache--broadcast pairs form a recovery star.
The pairs $(Z_2,B_i)$ recover $W_i$ for $i=3,\ldots,N$, and $(Z_1,C)$ recovers
$W_2$.  Conditioning on the cache of a user requesting $W_1$ and on that file
removes at least one file entropy from the broadcast.  Define the
\emph{delivery excesses}
\[
 \alpha_i\triangleq H(A_i)-F,
 \qquad
 \beta_i\triangleq H(B_i)-F,
 \qquad
 \gamma\triangleq H(C)-F.
\]

For any $X\in\{A_2,\ldots,A_N,B_3,\ldots,B_N,C\}$, choose a user $k$
who requests $W_1$ in that demand.  Cache--library independence and zero-error decoding give
\[
 H(X\mid Z_k)
 =H(W_1,X\mid Z_k)
 =F+H(X\mid W_1,Z_k).
\]
Consequently,
\begin{equation}
 H(X\mid W_1,Z_k)
 =H(X\mid Z_k)-F
 \le H(X)-F.
\label{eq:delivery-excess}
\end{equation}
Thus each excess is nonnegative.  Once $(W_1,Z_k)$ is known, appending
$X$ adds at most its delivery excess, since further conditioning reduces
entropy.
Let
\[
 \mathcal A\triangleq(A_2,\ldots,A_N).
\]

\begin{lemma}[Star Recovery]
\label{lem:star}
The broadcasts in $\mathcal A$ satisfy
\begin{equation}
 H(W_1,\mathcal A)\ge NF+\alpha_2.
\label{eq:star}
\end{equation}
\end{lemma}

\begin{proof}
Apply submodularity to the two collections
\[
 (W_1,\mathcal A)
 \quad\text{and}\quad
 (W_2,Z_3,A_2).
\]
Their intersection is $A_2$, and their union is $(W_1,W_2,Z_3,\mathcal A)$.  Hence
\begin{align*}
 &H(W_1,\mathcal A)+H(W_2,Z_3,A_2)\\
 &\quad\ge H(A_2)
   +H(W_1,W_2,Z_3,\mathcal A).
\end{align*}
For every $i=2,\ldots,N$, user $3$ recovers $W_i$ from $(Z_3,A_i)$.
The union therefore determines the entire library, so
\[
 H(W_1,W_2,Z_3,\mathcal A)
 \ge H(W_{[N]},Z_3,A_2).
\]
For demand $(1,1,2)$, user~3's decoding and security condition give
\[
 W_{[N]\setminus\{2\}}
 \ind (W_2,Z_3,A_2).
\]
Consequently,
\[
 H(W_{[N]},Z_3,A_2)
  =(N-1)F+H(W_2,Z_3,A_2).
\]
Substitution cancels the occurrence of $H(W_2,Z_3,A_2)$ on the left and yields
\[
 H(W_1,\mathcal A)
 \ge (N-1)F+H(A_2)
 =NF+\alpha_2.
\]
\end{proof}

\subsection{The Pivot Inequality}

To combine the recovery-star bound with the delivery-excess bounds, we
retain the exact conditional entropy of $B_3$ after $(W_1,Z_1)$ is observed.
Define this \emph{pivot residual} by
\[
 \rho\triangleq H(B_3\mid W_1,Z_1).
\]
The two tuples in the submodularity comparison share $(W_1,\mathcal A)$.
One also contains $Z_1$ and $B_3,\ldots,B_N$.  The other contains $Z_2$ and
$C$.  Security contributes $\rho$ to a lower bound for their union, exactly
canceling its contribution to the upper bound for the first tuple.

\begin{lemma}[Pivot Inequality]
\label{lem:pivot}
The delivery excesses obey
\begin{equation}
 H(Z_2)+\alpha_2
 +2\sum_{i=3}^{N}\alpha_i
 +\sum_{i=4}^{N}\beta_i+\gamma
 \ge (2N-2)F.
\label{eq:pivot-budget}
\end{equation}
\end{lemma}

\begin{proof}
Append $B_3$ to $(W_1,Z_1)$ first, retaining the exact entropy increment
$\rho$.  Then append all $A_i$ and $B_4,\ldots,B_N$.  User~1 requests $W_1$
in each of these demands.  Equation~\eqref{eq:delivery-excess} and further
conditioning therefore bound the remaining increments:
\begin{align}
 &H(W_1,Z_1,\mathcal A,B_3,\ldots,B_N)
 \nonumber\\
 &\qquad\le H(W_1,Z_1)+\rho
   +\sum_{i=2}^{N}\alpha_i
   +\sum_{i=4}^{N}\beta_i.
\label{eq:append1}
\end{align}
User $2$ requests $W_1$ in every $A_i$ and in $C$, so the same argument with
$(W_1,Z_2)$ gives
\begin{equation}
 H(W_1,Z_2,\mathcal A,C)
 \le H(W_1,Z_2)
 +\sum_{i=2}^{N}\alpha_i+\gamma.
\label{eq:append2}
\end{equation}

The two enlarged tuples have intersection $(W_1,\mathcal A)$ and union
\[
 U\triangleq
 (W_1,Z_1,Z_2,\mathcal A,B_3,\ldots,B_N,C).
\]
Submodularity bounds their entropy sum from below.  Combining this bound
with \eqref{eq:append1}--\eqref{eq:append2} gives
\begin{align}
 &H(W_1,Z_1)+H(W_1,Z_2)
 \nonumber\\
 &\quad\ge H(W_1,\mathcal A)+H(U)-\rho
 -2\sum_{i=2}^{N}\alpha_i
 -\sum_{i=4}^{N}\beta_i-\gamma.
\label{eq:pivot-submod}
\end{align}

The union $U$ contains $W_1$.  It also contains $(Z_1,C)$, which recovers
$W_2$, and $(Z_2,B_i)$, which recovers $W_i$ for every $i=3,\ldots,N$.
Thus $U$ determines the whole library, giving
\[
 H(U)\ge H(W_{[N]},Z_1,B_3).
\]
For the pivot delivery $B_3=X_{(1,3,1)}$, user $1$ requests $W_1$.
Decoding and the security condition for that demand give
\[
 W_{2:N}\ind(W_1,Z_1,B_3).
\]
It follows that
\begin{align*}
 H(W_{[N]},Z_1,B_3)
 &= (N-1)F+H(W_1,Z_1,B_3)\\
 &= NF+H(Z_1)+\rho.
\end{align*}
The last equality uses cache--library independence.  The $+\rho$ in
this lower bound cancels the $-\rho$ in \eqref{eq:pivot-submod}.

For $k=1,2$, cache--library independence also gives
\[
 H(W_1,Z_k)=F+H(Z_k).
\]
Substituting these cache--file identities, \eqref{eq:star}, and the lower
bound on $H(U)$ into \eqref{eq:pivot-submod} gives
\begin{align*}
 &2F+H(Z_1)+H(Z_2)\\
 &\quad\ge (NF+\alpha_2)
 +(NF+H(Z_1)+\rho)-\rho\\
 &\quad-2\sum_{i=2}^{N}\alpha_i
 -\sum_{i=4}^{N}\beta_i-\gamma.
\end{align*}
Canceling $H(Z_1)$ and rearranging gives exactly \eqref{eq:pivot-budget}.

\end{proof}

For a code with cache entropy at most $MF$ and delivery entropy at most $RF$,
each excess is at most $(R-1)F$.  The excesses in \eqref{eq:pivot-budget}
have total weight
\[
 1+2(N-2)+(N-3)+1=3N-5.
\]
The cancellation of $\rho$ removes the contribution of $B_3$ from this count.
Applying the resource bounds to \eqref{eq:pivot-budget} gives
\[
 MF+(3N-5)(R-1)F\ge(2N-2)F,
\]
and hence the high-memory supporting line
\[
 M+(3N-5)R\ge5N-7.
\]
Applying the finite-code inequality with resources $M+\varepsilon$ and
$R+\varepsilon$ along an achievable sequence and letting $\varepsilon\downarrow0$
gives the same supporting line for every achievable pair.

\subsection{Completion of the Exact Tradeoff}

For $N\ge4$, the converse in \cref{sec:low} gives
\[
 3M+(N-1)R\ge3N,
 \qquad\text{or}\qquad
 R\ge\frac{3(N-M)}{N-1},
\]
and \cref{sec:high} gives
\[
 R\ge\frac{5N-7-M}{3N-5}.
\]
The universal perfect-share bound gives $M\ge1$ and $R\ge1$.  The two nontrivial lines meet at
\[
 (M,R)=\left(\frac{N+1}{2},\frac32\right).
\]
The low-memory line meets $M=1$ at $(1,3)$, and the high-memory line meets $R=1$
at $(2N-2,1)$.  The constructions in \cite{fang2026secure}
achieve the three intersection points.  Memory sharing fills the two boundary
segments.  The code at $(2N-2,1)$ remains feasible for every larger memory
budget.
It follows that
\begin{equation}
 \begin{aligned}
 &R^\star_{N,3}(M)\\
 &\quad=\max\left\{
 \frac{3(N-M)}{N-1},
 \frac{5N-7-M}{3N-5},
 1
 \right\},\qquad M\ge1.
 \end{aligned}
\label{eq:boundary-general}
\end{equation}

For $N=3$, \cref{sec:small} gives the two bounds
\[
 2M+R\ge5,
 \qquad
 M+R\ge\frac72.
\]
The bound from \cref{sec:high} becomes
\[
 M+4R\ge8,
\]
and the share bound gives $R\ge1$.  The four boundary corners are
\[
 (1,3),\qquad
 \left(\frac32,2\right),\qquad
 \left(2,\frac32\right),\qquad
 (4,1).
\]
Specializing the constructions in \cite{fang2026secure} to $N=3$
achieves $(1,3)$, $(2,3/2)$, and $(4,1)$.  The closed-form $\mathbb F_3$
code in \cref{thm:n3closed} achieves $(3/2,2)$.  Memory sharing fills the three
boundary segments.  The code at $(4,1)$ remains feasible for every larger
memory budget.  Hence
\begin{equation}
 \begin{aligned}
 &R^\star_{3,3}(M)\\
 &\quad=\max\left\{
 5-2M,\frac72-M,2-\frac M4,1
 \right\},\qquad M\ge1.
 \end{aligned}
\label{eq:boundary-three}
\end{equation}

\begin{remark}[Two-File Baseline]\label{rem:n2baseline}
For $N=2$, set $d=H(W_2\mid W_1,Z_1,Z_2)$ under demand $(1,1,2)$.  The two-cache residual bound and \eqref{eq:completion-two-file} give
\[
 H(Z_2)+d\ge2F,
 \qquad
 H(X_{112})\ge F+d.
\]
Hence
\[
 H(Z_2)+H(X_{112})\ge3F,
 \qquad
 M+R\ge3.
\]
Together with $R\ge1$, the endpoint constructions at $(1,2)$ and $(2,1)$
from \cite{fang2026secure} and their memory-sharing segment meet
these bounds.  The code at $(2,1)$ remains feasible for every larger memory
budget.
\end{remark}

The two-file result in \cref{rem:n2baseline} is
\begin{equation}
 R^\star_{2,3}(M)=\max\{3-M,1\},\qquad M\ge1.
\label{eq:boundary-two}
\end{equation}
Together with the universal condition $M\ge1$, these formulas establish \cref{thm:main} for every $N\ge2$.

\section{Conclusion}
\label{sec:conclusion}

We have determined the exact zero-error memory--rate tradeoff for secure
coded caching with three users and every library size $N\ge2$.  For $N\ge4$,
the two converse bounds meet the known achievable corners.  For $N=3$, our
explicit $\mathbb F_3$ code achieves the additional corner $(M,R)=(3/2,2)$
required by the converse.
The low-memory converse follows from overlapping recoveries under a repeated
demand and a comparison of requested and unrequested file residuals.  The
same placement must also serve distinct demands, which bounds the joint
saving in broadcast entropy.  At the new three-file corner,
the symmetrized equality conditions $p=0,q=F$ specify the information available
to pairs and triples of caches.  Our block-MDS placement realizes this
information pattern.  Each user recovers its requested file in the quotient
by the key image.  Its cache and the broadcast jointly reveal no information
about the complete tuple of unrequested files.

\appendices
\section{Closed-Form Coordinates for the Three-File Code}
\label{app:n3-closed-form}
The coordinate formulas below verify the support-selection property
\eqref{eq:kappa-selector} and specify every broadcast in \cref{sec:n3ach}.
All arithmetic is over $\mathbb F_3$.

\subsection*{The Placement Matrix}

Stacking the syndrome rows $r_{12},r_{13},r_{23},r_{123}$ from
\eqref{eq:n3placement} gives
\[
\Theta=
\begin{bmatrix}
0&2&2&0&2&1\\
2&0&1&1&1&1\\
2&1&0&1&1&0\\
0&2&2&1&0&2
\end{bmatrix}.
\]
The matrix $\Theta$ has rank four.  The $4\times4$ submatrices retaining
file-block pairs $12,13,23$ have determinants $2,1,2$, respectively.
The kernel of $\Theta$ is
\[
\ker\Theta
=\operatorname{row}\begin{bmatrix}
1&1&1&2&1&0\\
0&1&0&2&0&1
\end{bmatrix}.
\]
For each file block, the restriction of the corresponding coordinate
projection to $\ker\Theta$ is an isomorphism onto that block.  Hence separate
changes of basis within the files turn the
kernel into the diagonal subspace
$\{(u,u,u):u\in\mathbb F_3^2\}$.  Together with invertible row operations,
these changes put $\Theta$ in the form \eqref{eq:blockmds}.

\subsection*{Coordinates for the Three Cancellation Planes}

We first choose coordinates for any hyperplane with a normalized normal
$\ell=(1,\ell_2,\ell_3,\ell_4,\ell_5)\in\mathbb F_3^5$.  Define the
$4\times5$ matrix
\[
V(\ell)=
\begin{bmatrix}
-\ell_2&1&0&0&0\\
-\ell_3&0&1&0&0\\
-\ell_4&0&0&1&0\\
-\ell_5&0&0&0&1
\end{bmatrix}.
\]
The rows $(v_1,\ldots,v_4)$ of $V(\ell)$ form a basis of $\ker\ell$.

Now fix one of the four supports $J\in\{12,13,23,123\}$ and use the normal
$\ell_J$ from \eqref{eq:support-hyperplanes}.  Define the fixed
$2\times2$ matrix $A_J$ by
\[
\begin{array}{c|c@{\qquad}c|c}
J&A_J&J&A_J\\ \hline
12&\begin{bmatrix}1&0\\2&1\end{bmatrix}&
13&\begin{bmatrix}2&0\\2&1\end{bmatrix}\\[3mm]
23&\begin{bmatrix}2&0\\0&1\end{bmatrix}&
123&I_2.
\end{array}
\]
Write $H=H_J$ and $D_k=D_k^J=C_k\cap H_J$ for $k=1,2,3$.
In the basis given by $V(\ell_J)$, the cancellation planes have coordinates
\begin{align*}
D_1&=\operatorname{row}[I_2\mid0],&
D_2&=\operatorname{row}[0\mid I_2],\\
D_3&=\operatorname{row}[A_J\mid I_2].
\end{align*}
Every $A_J$ is invertible.  Thus the projections
$\pi_1:D_3\to D_1$ and $\pi_2:D_3\to D_2$ are isomorphisms.
Their coordinate matrices are $A_J$ and $I_2$, respectively.

\subsection*{The Extension-Defect Blocks}

To check which file blocks remain in the extension defect, express each
map $\kappa_{J,j}:D_3^J\to\mathcal V_j^*$ in coordinates.
Use the two-row basis $[A_J\mid I_2]$ for $D_3^J$ and the standard
coefficient coordinates in the dual file space $\mathcal V_j^*$.
The three $2\times2$ block matrices are
\[
\begin{array}{c|c|c|c}
J&\kappa_{J,1}&\kappa_{J,2}&\kappa_{J,3}\\ \hline
12&
\begin{bmatrix}1&2\\1&0\end{bmatrix}&
\begin{bmatrix}2&2\\2&0\end{bmatrix}&
\begin{bmatrix}0&0\\0&0\end{bmatrix}\\[3mm]
13&
\begin{bmatrix}0&2\\1&1\end{bmatrix}&
\begin{bmatrix}0&0\\0&0\end{bmatrix}&
\begin{bmatrix}1&1\\1&2\end{bmatrix}\\[3mm]
23&
\begin{bmatrix}0&0\\0&0\end{bmatrix}&
\begin{bmatrix}1&0\\0&1\end{bmatrix}&
\begin{bmatrix}2&0\\1&1\end{bmatrix}\\[3mm]
123&
\begin{bmatrix}1&0\\0&2\end{bmatrix}&
\begin{bmatrix}0&2\\2&1\end{bmatrix}&
\begin{bmatrix}1&2\\0&2\end{bmatrix}.
\end{array}
\]
The determinant triples, in the row order of the table, are
$(1,2,0)$, $(1,0,1)$, $(0,1,2)$, and $(2,2,2)$.
Thus the blocks outside $J$ vanish and the blocks on $J$ are invertible,
proving \eqref{eq:kappa-selector}.

Requiring the unrequested block to vanish determines each two-file
support normal.  Let
$\ell=(1,a,b,c,d)$ vary over normalized normals in $\mathbb F_3^5$,
and require the planes $C_k\cap\ker\ell$ to be pairwise complementary
in $\ker\ell$.  The pairwise cache intersections are
$\langle e_1\rangle$, $\langle e_2\rangle$, and $\langle e_4\rangle$.
The normal $\ell$ must be nonzero on each of these one-dimensional spaces.
Since its first coordinate is one, the requirement is exactly $a,c\ne0$.
Define the determinant polynomials
\begin{align*}
\delta_1&=ad+a+bc-b-c-d,\\
\delta_2&=-ac+ad+a+bc+b+c+d,\\
\delta_3&=ac+ad-a-bc-b+c-d.
\end{align*}
For this normal $\ell$, define $\kappa_\ell$ by the same
difference of file-label maps as in \eqref{eq:kappa}, using the
decomposition of $\ker\ell$ into $C_1\cap\ker\ell$ and
$C_2\cap\ker\ell$.
To compute its block determinants, write a vector $h\in C_3\cap\ker\ell$
using scalars $x,y,z\in\mathbb F_3$:
\begin{align*}
h&=xe_2+ye_4+z(e_3+e_5),\\
ax+cy+(b+d)z&=0.
\end{align*}
The extension defect is
\[
\kappa_\ell(h)
=xr_{13}+yr_{23}+zr_{123}+(cy+dz)r_{12}.
\]
Because $a\ne0$, the constraint determines $x$ from $y,z$.
Let $K^{(y)}$ and $K^{(z)}$ be the rows obtained by evaluating
$\kappa_\ell$ at the basis vectors specified by
$(y,z)=(1,0)$ and $(0,1)$, respectively.  Substitution gives
\begin{align*}
 K^{(y)}&=-\frac ca r_{13}+r_{23}+c r_{12},\\
 K^{(z)}&=-\frac{b+d}{a}r_{13}+r_{123}+d r_{12}.
\end{align*}
For $j=1,2,3$, write $K^{(y)}_j,K^{(z)}_j$ for the two entries in
file block $j$ of the corresponding row.  In the $(y,z)$ basis, the
block determinant is
\[
 \det\begin{bmatrix}K^{(y)}_j\\K^{(z)}_j\end{bmatrix}
 =\frac{\delta_j}{a}.
\]
Since $a\ne0$, each block has rank two exactly when $\delta_j\ne0$.

An off-support block must vanish in both rows.  These equations also
explain the choice of each two-file normal.  For off-support block $j=3$,
the two coordinates of $aK^{(y)}_3=0$ are
$a-c+2ac=0$ and $c(a-1)=0$.  Since $a,c\ne0$, they give
$a=1$, $c=2$.  Substituting into $aK^{(z)}_3=0$ gives
$b=d=2$.  For $j=2$, the equations $c(2a-1)=0$ and $a-c=0$
give $a=c=2$, and the second row then gives $b=d=1$.
For $j=1$, the equations $2(a-c)=0$ and $a(1+2c)=0$
give $a=c=1$, followed by $d=2$ and $b=1$.
Hence the zero-block equations uniquely determine
$\ell_{12},\ell_{13},\ell_{23}$ in \eqref{eq:support-hyperplanes}.
The nonzero determinants above verify the requested-block ranks for these
three normals and for $\ell_{123}$.

\subsection*{The Closed-Form Delivery Matrix}

For $j\in\{1,2,3\}$ and a $2\times2$ matrix $M$, let
$\mathcal E_j(M)$ denote the $2\times6$ matrix whose $j$th two-column
block is $M$ and whose other
blocks are zero.  Fix a nonconstant demand $\bm d$ with requested file
set $J=\{d_1,d_2,d_3\}$, and choose $\alpha_1,\alpha_2$ by
\eqref{eq:alpha-rule}.  Define the two $2\times2$ matrices
\begin{align*}
\Theta_1&=\alpha_1A_J^{-1}\kappa_{J,d_1},&
\Theta_2&=\alpha_2\kappa_{J,d_2}.
\end{align*}
The $4\times6$ file matrix is
\begin{equation}
B_{\bm d}=
\begin{bmatrix}
-\mathcal E_{d_1}(\Theta_1)\\[1mm]
-\ell_{J,4}r_{12}\\
-\ell_{J,5}r_{12}
\end{bmatrix}
-
\begin{bmatrix}
0_{2\times6}\\[1mm]
\mathcal E_{d_2}(\Theta_2)
\end{bmatrix}.
\label{eq:explicit-BV}
\end{equation}
The key matrix is $V_{\bm d}=V(\ell_J)$.
Equation \eqref{eq:explicit-BV} is the coordinate form of
\eqref{eq:b-interpolation}.  The factor $A_J^{-1}$ represents
$\pi_1^{-1}$.  The file-label map $\varphi_2$ contributes
$-\ell_{J,4}r_{12}$ and $-\ell_{J,5}r_{12}$ in rows 3 and 4,
respectively.

\section{Equality Structure of the Low-Memory Bound}
\label{app:exchange-audit}
This appendix characterizes equality on the general low-memory face.
The four-core inequality supplies an additional constraint on optimal
placement. Combining it with the main-text bounds yields the refined
entropy identity and the equality conditions below.

\subsection{The Four-Core Bound}

For this subsection fix $N\ge4$ and a finite symmetric code. Set
$X=X_{(A,A,B)}$, $\bZ=(Z_1,Z_2,Z_3)$, and
$\mathcal L=(Z_1,Z_2,Z_3,X)$.
Two overlapping observations recover $A,B$ and constrain the residual
broadcast entropy in terms of cache entropies. The bound counts the
broadcast entropy twice. In this subsection the library is $(A,B,C,D,Q)$,
where $Q$ contains $m=N-4$ files and is empty for $N=4$.

\begin{lemma}[Four-Core Cycle Bound]
\label{lem:fourcore}
For the library $(A,B,C,D,Q)$, with $Q$ containing $m=N-4$ files, and demand $(A,A,B)$,
\begin{equation}
 2H(Z_1)+2H(X)
 \ge8F+I(Q;\bZ\mid A,B,C).
\label{eq:fourcore-synergy}
\end{equation}
Equivalently,
\[
 2H(Z_1)+2H(X)+H(Q\mid A,B,C,\bZ)
 \ge(m+8)F.
\]
\end{lemma}

\begin{proof}
Put $z=H(Z_1)=H(Z_2)=H(Z_3)$ and $x=H(X)$. Define the conditional cache
entropy $v$, residual broadcast entropy $u$, and cache--broadcast
mutual information $i$ by
\begin{align*}
 v&=H(Z_1,Z_2\mid A)=H(Z_1,Z_3\mid B),\\
 u&=H(X\mid A,Z_1,Z_2),\qquad i=I(Z_2;X).
\end{align*}
The equality defining $v$ uses placement symmetry.
The two users requesting $A$ give, by same-secret uncrossing,
\begin{equation}
 2z+x\ge2F+v+u+i.
 \label{eq:four-common-recovery}
\end{equation}
Now compare $(Z_1,Z_3,X)$ and $(Z_2,Z_3,X)$.  Each recovers $A,B$.
Their common observation $(Z_3,X)$ is independent of $A$, so
submodularity conditioned on $A$ gives
\begin{align*}
 H(Z_1,Z_3,X)+H(Z_2,Z_3,X)
 \ge H(\mathcal L)+H(Z_3,X)+F.
\end{align*}
To bound the two pair entropies, put $i_3=I(Z_3;X)$.  Recovery of $B$
from $(Z_3,X)$ gives
\begin{align*}
 H(Z_1,Z_3,X)
 &=H(Z_3,X)+H(Z_1\mid B,Z_3,X)\\
 &\le z+x-i_3+H(Z_1\mid B,Z_3)
 =v+x-i_3.
\end{align*}
The same bound holds with $Z_2$ in place of $Z_1$.

Define the conditional cache entropies
\begin{align*}
 h_2&=H(\bZ\mid A,C),\\
 h_3&=H(\bZ\mid A,B,C),\\
 h_-&=H(\bZ\mid A,C,D,Q).
\end{align*}
Because $\mathcal L$ recovers $A,B$, placement symmetry gives
$H(\mathcal L)\ge2F+h_2$.  Substituting the pair bounds and
$H(Z_3,X)=z+x-i_3$ into the overlap inequality yields
\begin{equation}
 2v-z+x\ge3F+h_2.
 \label{eq:four-recovery-overlap}
\end{equation}

Given $A,C$ and all caches, the broadcast still recovers $B$.  Thus
\begin{equation}
 u\ge H(B\mid A,C,\bZ)=F+h_3-h_2.
 \label{eq:four-file-residual}
\end{equation}
User 2's joint security provides a second constraint:
\begin{align*}
 H(Z_2,X)
 &=H(Z_2,X\mid B,C,D,Q)\\
 &\le H(\mathcal L\mid B,C,D,Q)
 =H(\mathcal L\mid A,C,D,Q)\\
 &\le h_-+u.
\end{align*}
The middle equality holds because $\mathcal L$ recovers both $A,B$,
which are independent of $(C,D,Q)$.  The last inequality follows by
expanding the conditional entropy and then dropping conditioning
variables from the broadcast term.  Since $H(Z_2,X)=z+x-i$,
\begin{equation}
 u\ge z+x-i-h_-.
 \label{eq:four-private-observation}
\end{equation}

Twice \eqref{eq:four-common-recovery}, together with
\eqref{eq:four-recovery-overlap}, gives
\[
 3z+3x\ge7F+h_2+2u+2i.
\]
Adding \eqref{eq:four-file-residual} and
\eqref{eq:four-private-observation} supplies
\[
 2u\ge F+h_3-h_2+z+x-i-h_-.
\]
Substitution cancels $h_2$ and gives
\[
 2z+2x\ge8F+h_3-h_-+i.
\]
Placement symmetry exchanges $B,D$ in a cache entropy, so
\[
 h_3-h_-=I(Q;\bZ\mid A,B,C).
\]
Dropping $i\ge0$ proves \eqref{eq:fourcore-synergy}.
Section~\ref{app:core-slacks} gives the exact gaps of the four constraints.
\end{proof}

In the notation $z=H(Z_1)$, $x=H(X_{112})$, and $q$ of
\eqref{eq:pq}, the four-core bound gives
\begin{equation}
 2z+2x\ge8F+(N-4)q.
 \label{eq:budget-four}
\end{equation}
Its strengthened right-hand side includes the nonnegative correction
$\Gamma_{N-4}$ from \cref{lem:tail}.
For the four-file core, use $W_N$ as the fourth core file and let $Q$ be
the remaining tuple $W_{4:N-1}$.  Applying Lemma~\ref{lem:fourcore} with
\[
 (A,B,C,D,Q)=(W_1,W_2,W_3,W_N,W_{4:N-1}),
\]
together with \eqref{eq:tail-three}, gives
\[
 I(W_{4:N-1};\bZ\mid W_{1:3})
 =(N-4)q+\Gamma_{N-4},
\]
and hence \eqref{eq:budget-four}, including its strengthened version.

\subsection{A Refined Entropy Identity}

Fix $N\ge4$ and a finite symmetric code. Use $z,x,y,p,q$ from
Section~\ref{sec:low}, so $y=H(X_{123})$. To identify the equality
conditions, retain the two delivery entropies in
the functional $3z+2x+(N-3)y-3NF$ and write it as a sum of nonnegative gaps.
Let $G_4$ be the gap in \eqref{eq:budget-four}, let $G_2,G_3$ be
the gaps in \cref{lem:synergy-cone}, and let $G_D$ be the gap in
\eqref{eq:distinct}:
\begin{align*}
G_4&=2z+2x-8F-(N-4)q,\\
G_2&=z-F-(N-1)p,\\
G_3&=2z+x-5F-p-(N-3)q,\\
G_D&=y-3F+p+q.
\end{align*}
For $N\ge4$, define the weights
\begin{equation}
\lambda_4=\frac{N-3}{N-2},\qquad
\lambda_2=\frac{N-4}{N-2},\qquad
\lambda_3=\frac{2}{N-2}.
\label{eq:lambdas}
\end{equation}
Matching the coefficients of $z,x,y$ and canceling $p,q$ gives
\eqref{eq:lambdas}.  Expanding the gaps gives
\begin{equation}
\begin{aligned}
&3z+2x+(N-3)y-3NF\\
&\quad=\lambda_4G_4+\lambda_2G_2+\lambda_3G_3\\
&\qquad +(N-3)G_D.
\end{aligned}
\label{eq:gap-balance}
\end{equation}

\begin{remark}[The Four-File Threshold and the Three-File Case]
\label{rem:n3sign}
The coefficient $\lambda_2=(N-4)/(N-2)$ in \eqref{eq:lambdas} is positive for
$N>4$, zero at $N=4$, and negative at $N=3$.  For three files, the identity
$p+(F-p)=F$ eliminates the placement information between the three-core and
delivery-completion inequalities, yielding the face $M+R=7/2$.  Its
intersection with the three-core face is $(3/2,2)$, where the equality
conditions $p=0,q=F$ are realized by the code in \cref{sec:n3ach}.
\end{remark}

For $N\ge4$, all coefficients on the right of
\eqref{eq:gap-balance} are nonnegative, giving
\[
 3H(Z_1)+2H(X_{112})+(N-3)H(X_{123})\ge3NF.
\]
On the supporting line, the alphabet upper bounds and every gap with
positive weight must be tight.

\subsection{Equality Conditions}

For a finite symmetric code, define its actual normalized alphabet sizes by
\begin{align*}
M_{\mathrm a}&=\max_{k\in[3]}\frac{\log_2|\mathcal Z_k|}{F},\\
R_{\mathrm a}&=\max_{\bm d\in[N]^3}\frac{\log_2|\mathcal X_{\bm d}|}{F}.
\end{align*}
Equality on the low-memory face forces the active cache and broadcasts
to attain these alphabet bounds.

\begin{corollary}[Rigidity on the Low-Memory Face]\label{cor:rigidity}
For $N\ge4$, suppose a finite symmetric code has normalized alphabet sizes
$(M_{\mathrm a},R_{\mathrm a})$ on the face
\[
3M_{\mathrm a}+(N-1)R_{\mathrm a}=3N.
\]
The active cache and delivery variables saturate their alphabets:
\begin{equation}
 H(Z_1)=M_{\mathrm a}F,\qquad
 H(X_{112})=H(X_{123})=R_{\mathrm a}F.
\label{eq:entropy-saturation-low}
\end{equation}
For $N>4$,
\[
 p=\frac{M_{\mathrm a}-1}{N-1}F,
 \qquad
 q=\frac{2(M_{\mathrm a}-1)}{N-1}F=2p,
\]
and all four gaps in \eqref{eq:gap-balance}, together with the tail correction terms
in \cref{lem:synergy-cone} and \eqref{eq:budget-four}, vanish.  At the interior corner
$M_{\mathrm a}=(N+1)/2$,
\[
 p=\frac F2,\qquad q=F,
\]
so the three caches, together with any two anchor files, determine the rest of
the library.
\end{corollary}

\begin{proof}
The entropy upper bounds and the face equation give
\[
3z+2x+(N-3)y-3NF
\le
\bigl(3M_{\mathrm a}+(N-1)R_{\mathrm a}-3N\bigr)F=0.
\]
Equation~\eqref{eq:gap-balance} gives the reverse inequality.  The three active
entropy--alphabet chains are
\begin{align*}
 z&\le\log_2|\mathcal Z_1|\le M_{\mathrm a}F,\\
 x&\le\log_2|\mathcal X_{112}|\le R_{\mathrm a}F,\\
 y&\le\log_2|\mathcal X_{123}|\le R_{\mathrm a}F.
\end{align*}
Their coefficients $3$, $2$, and $N-3$ are positive, so equality forces every
inequality in these chains to be tight, proving
\eqref{eq:entropy-saturation-low}.

For $N>4$, every coefficient on the right of \eqref{eq:gap-balance} is
strictly positive.  Hence $G_4=G_2=G_3=G_D=0$.  The strengthened forms of
\cref{lem:synergy-cone} and \eqref{eq:budget-four} force all three
tail-correlation slacks to vanish as well.
Solving $G_2=G_D=0$ together with the face equation yields the displayed values
of $p$ and $q$.  At $M_{\mathrm a}=(N+1)/2$, the equality $q=F$, file symmetry,
and the vanishing tail slack give
\[
 H(W_{3:N}\mid W_1,W_2,\bZ)=0.
\]

\end{proof}

\begin{remark}[Boundary Profile at Four Files]
For a finite symmetric code on the $N=4$ low-memory face, the coefficient
$\lambda_2=0$ removes $G_2$ from the weighted sum in \eqref{eq:gap-balance}.
Equality on the face forces $G_4=G_3=G_D=0$, and the strengthened three-core
inequality further forces $\Gamma_1=0$:
\begin{equation}
 G_4=G_3=G_D=0,
 \qquad \Gamma_1=0,
\label{eq:n4-active-equalities}
\end{equation}
These equalities fix the sum of the two placement coordinates and constrain
their individual values:
\[
 p+q=(M_{\mathrm a}-1)F,
 \qquad
 \max\{0,M_{\mathrm a}-2\}F
 \le p\le\frac{M_{\mathrm a}-1}{3}F.
\]
Indeed, the strengthened three-core inequality
\[
2z+x\ge5F+p+q+\Gamma_1
\]
and $G_3=0$ force $\Gamma_1=0$.
The equations $G_3=G_D=0$, together with
$M_{\mathrm a}+R_{\mathrm a}=4$, determine the sum $p+q$.
The inequalities $G_2\ge0$ and $0\le p,q\le F$ give the stated necessary
interval for $p$.  At the two endpoints of the low-memory face, the interval
collapses to $(p,q)=(0,0)$ at $M_{\mathrm a}=1$ and $(p,q)=(F/2,F)$ at
$M_{\mathrm a}=5/2$.

\end{remark}

\begin{remark}[Linear-Rank Interpretation of the Placement Coordinates]\label{rem:linear-rank}
Let $s$ be a prime power.  Let $U,V,T$ be random vectors obtained by linear
maps over $\mathbb F_s$ from a common uniform base vector, and let
$\mathcal U,\mathcal V,\mathcal T$ be the row spaces of the respective maps.
Normalize the base-two entropy and mutual information by one field symbol:
\[
 H_s(U\mid T):=\frac{H(U\mid T)}{\log_2 s},
 \qquad
 I_s(U;V\mid T):=\frac{I(U;V\mid T)}{\log_2 s}.
\]
Then
\[
 H_s(U\mid T)=\dim\frac{\mathcal U+\mathcal T}{\mathcal T},
\]
and subspace modularity gives
\[
 I_s(U;V\mid T)
 =\dim\bigl((\mathcal U+\mathcal T)\cap
             (\mathcal V+\mathcal T)\bigr)-\dim\mathcal T.
\]
Thus, in a linear realization, $p/\log_2s$ and $q/\log_2s$ are
intersection-space dimensions in successive quotient spaces.  Every
conditional mutual information in a requester-cycle gap becomes the
corresponding rank-submodularity slack, equivalently an excess intersection
dimension in the quotient.  For a finite symmetric linear code on the $N>4$
low-memory equality face, tightness makes all four gaps vanish and fixes the
two intersection dimensions.

The block-MDS quotient in \cref{sec:n3ach} realizes $p=0,q=F$, whereas any
finite symmetric linear code attaining the general $N>4$ interior corner
has $p=F/2,q=F$.  Let $f:=F/\log_2s$ be the file length in field symbols.
In the three-file profile, the two-cache intersection associated with $p$
has dimension zero.  At the general interior corner, it has dimension $f/2$.
The three-cache intersection associated with $q$ has dimension $f$ in both
profiles.
\end{remark}

\subsection{Exact Slacks in the Core Bounds}
\label{app:core-slacks}

Equality in \cref{lem:threecore,lem:fourcore} is determined by
the gaps in their chain-rule and submodularity steps.  Each gap below is
a sum of conditional entropies and mutual informations.

\subsubsection{Three-Core Slacks}

For $N\ge3$, use the library $(A,B,C,Q)$, where $Q$ contains $m=N-3$ files, and the
fixed demand $(A,A,B)$.  Let $e,w,\mathcal Y,\mathcal L$ be as in
\cref{lem:repeated-view}.  Retaining the slacks in the two submodularity
comparisons and the final secret peeling gives
\begin{align}
 H(Z_2,X)-3F-e-w
 ={}&I(Z_1;Z_2\mid A,X)\nonumber\\
 &+I(Z_2;Z_3\mid B,Z_1,X)\nonumber\\
 &+I(Z_3;X)+I(Z_1;X\mid B,Z_3).
 \label{eq:repeated-view-gap}
\end{align}
The file-residual comparison has the exact slack
\begin{align}
 e-H(C\mid\mathcal L)
 ={}&H(X\mid A,B,\bZ)\nonumber\\
 &+I(C,Z_3;X\mid A,Z_1,Z_2).
 \label{eq:file-residual-gap}
\end{align}
In \cref{lem:file-residual}, the first and last conditioning steps
contribute $I(X;Z_3\mid A,Z_1,Z_2)$ and $I(C;X\mid A,\bZ)$, whose sum
is the mutual information above.  The decoding step contributes
$H(X\mid A,B,\bZ)$.

Define the groups
\begin{align*}
\mathsf L_3&=H(X\mid A,B,\bZ)
 +I(Q;X\mid A,B,C,\bZ)\\
&\quad+H(Z_1,Z_3\mid B,C,Q,Z_2,X),\\
\mathsf P_3&=I(Z_3;X)+I(Z_1;X\mid B,Z_3),\\
\mathsf H_3&=I(C,Z_3;X\mid A,Z_1,Z_2),\\
\mathsf S_3&=I(Z_1;Z_2\mid A,X)
 +I(Z_2;Z_3\mid B,Z_1,X).
\end{align*}
The accounting identity \eqref{eq:core-cut-accounting} and the two
slack identities give
\begin{align}
 &H(A,B,Z_1,Z_2,X)-5F-K_3\nonumber\\
 &\quad=H(X\mid A,B,\bZ)+\mathsf P_3+\mathsf H_3+\mathsf S_3
 \ge0.
 \label{eq:threecore-hidden-budget}
\end{align}
The remaining-file information has the exact difference
\begin{align*}
 K_3-\sigma_Q
 ={}&H(Z_1,Z_3\mid B,C,Q,Z_2,X)\\
 &+I(Q;X\mid A,B,C,\bZ).
\end{align*}
Here user 2's joint security makes $Q$ independent of
$(A,B,C,Z_2,X)$, and $A$ is recovered from $(Z_2,X)$.
Adding this difference yields the full three-core gap.  Define
\[
 \Phi_3=H(A,B,Z_1,Z_2,X)
       +H(Q\mid A,B,C,\bZ)-(m+5)F.
\]
Then
\begin{equation}
 \Phi_3=\mathsf L_3+\mathsf P_3+\mathsf H_3+\mathsf S_3.
 \label{eq:threecore-gap}
\end{equation}
Repeated-request uncrossing further gives
\[
 2H(Z_1)+H(X)-5F-\sigma_B-\sigma_Q
 =\mathsf U_3+\mathsf L_3+\mathsf P_3+\mathsf H_3+\mathsf S_3.
\]

\subsubsection{Four-Core Slacks}

For $N\ge4$, use the four-core library $(A,B,C,D,Q)$ with $m=N-4$ remaining files.
Keep the quantities defined in the proof of \cref{lem:fourcore}.
The four constraints in that proof have gaps
\begin{align*}
 \eta_1&=2z+x-2F-v-u-i,\\
 \eta_2&=2v-z+x-3F-h_2,\\
 \eta_3&=u-F-h_3+h_2,\\
 \eta_4&=u+h_--z-x+i.
\end{align*}
Keeping the chain-rule and submodularity slacks gives
\begin{align}
 \eta_1={}&I(Z_1;X)+I(Z_1;Z_2\mid A,X),\nonumber\\
 \eta_2={}&I(Z_3;X)+I(Z_1;X\mid B,Z_3)\nonumber\\
 &+I(Z_2;X\mid B,Z_3)
  +I(Z_1;Z_2\mid A,Z_3,X)\nonumber\\
 &+H(X\mid A,B,\bZ),\nonumber\\
 \eta_3={}&I(C,Z_3;X\mid A,Z_1,Z_2)
          +H(X\mid A,B,C,\bZ),\nonumber\\
 \eta_4={}&I(C,D,Q,Z_3;X\mid A,Z_1,Z_2)\nonumber\\
 &+H(Z_1,Z_3\mid B,C,D,Q,Z_2,X).
 \label{eq:four-short-slacks}
\end{align}
For $\eta_2$, the terms come from the two pair-entropy upper bounds,
their overlap, and the entropy left after recovering $A,B$.
For $\eta_4$, security gives
$H(Z_2,X)=H(Z_2,X\mid B,C,D,Q)$.
Adding the other caches contributes the last conditional entropy in
$\eta_4$.  The change in the conditional broadcast entropy contributes
its mutual-information term.

Since $h_3-h_-=I(Q;\bZ\mid A,B,C)$, define
\[
 \Phi_4=2H(Z_1)+2H(X)-8F-I(Q;\bZ\mid A,B,C).
\]
The same combination as in the proof gives the exact identity
\begin{equation}
 \Phi_4=i+2\eta_1+\eta_2+\eta_3+\eta_4.
 \label{eq:four-short-gap}
\end{equation}
For use with the equality conditions, group the terms as
\begin{align*}
\mathsf L_4&=H(X\mid A,B,\bZ)+H(X\mid A,B,C,\bZ)\\
&\quad+H(Z_1,Z_3\mid B,C,D,Q,Z_2,X),\\
\mathsf P_4&=2I(Z_1;X)+I(Z_2;X)+I(Z_3;X),\\
\mathsf H_4&=2I(C,Z_3;X\mid A,Z_1,Z_2)\\
&\quad+I(D,Q;X\mid A,C,\bZ),\\
\mathsf E_4&=I(Z_1;X\mid B,Z_3)+I(Z_2;X\mid B,Z_3),\\
\mathsf S_4&=2I(Z_1;Z_2\mid A,X)
 +I(Z_1;Z_2\mid A,Z_3,X).
\end{align*}
The mutual-information chain rule splits the $C,D,Q,Z_3$ term in
$\eta_4$ into its $C,Z_3$ part and its $D,Q$ part.
Thus \eqref{eq:four-short-gap} becomes
\begin{equation}
 \Phi_4=\mathsf L_4+\mathsf P_4+\mathsf H_4+\mathsf E_4+\mathsf S_4.
 \label{eq:fourcore-gap}
\end{equation}
Thus equality in either core bound requires every component with positive
weight to vanish.

\section{A General Same-Secret Identity}
\label{app:general-uncrossing}
The two-share identity used in the converse extends to an arbitrary
number of shares and common side information. We retain the full
identity here, using the conditional total correlation defined in
Section~\ref{sec:entropy-identities}.

\begin{lemma}[Same-Secret Uncrossing for $r$ Shares]\label{lem:r-uncross}
Let $r$ be a nonnegative integer and put $\mathcal R=\{1,\ldots,r\}$.
Let $A,G,X,Z_1,\ldots,Z_r$ be finite-alphabet random variables, where
$G$ denotes common side information. Suppose that the same secret $A$
is recoverable from $(G,Z_i,X)$ for every $i\in\mathcal R$. Then
\begin{align}
&\sum_{i\in\mathcal R}H(Z_i\mid G)+H(X\mid G)\nonumber\\
={}&H(A\mid G)+(r-1)H(A\mid X,G)\nonumber\\
&+H(Z_{\mathcal R},X\mid A,G)\nonumber\\
&+\sum_{i\in\mathcal R}I(Z_i;X\mid G)\nonumber\\
&+\TC(Z_{\mathcal R}\mid A,X,G).
\label{eq:r-uncross}
\end{align}
\end{lemma}

\begin{proof}
For each $i$, recoverability gives
\[
  H(Z_i,X\mid G)
  =H(A\mid G)+H(Z_i,X\mid A,G).
\]
Also,
\[
  H(Z_i\mid G)+H(X\mid G)
  =H(Z_i,X\mid G)+I(Z_i;X\mid G).
\]
Sum the latter equality over $i$ and subtract $(r-1)H(X\mid G)$.  To combine the conditional entropies from the first equality, use the definition of total correlation:
\begin{align*}
  \sum_{i\in\mathcal R}H(Z_i,X\mid A,G)
  ={}&(r-1)H(X\mid A,G)\\
  &+H(Z_{\mathcal R},X\mid A,G)\\
  &+\TC(Z_{\mathcal R}\mid A,X,G).
\end{align*}
The remaining secret and delivery terms satisfy
\begin{align*}
  &rH(A\mid G)-(r-1)H(X\mid G)
    +(r-1)H(X\mid A,G)\\
  &\qquad=H(A\mid G)+(r-1)H(A\mid X,G),
\end{align*}
because $H(X\mid G)-H(X\mid A,G)=I(A;X\mid G)$.  Combining the displayed identities proves \eqref{eq:r-uncross}.
\end{proof}

\bibliographystyle{IEEEtran}
\bibliography{references}
\end{document}